\documentclass[11pt]{article}
\usepackage{jheppub}
\usepackage[utf8]{inputenc}
\usepackage{amsmath}
\usepackage{amssymb}
\usepackage{commath}
\usepackage[mathscr]{euscript}
\usepackage{color}
\usepackage{physics}
\usepackage{braket}
\usepackage{tensor}
\usepackage{hyperref}
\usepackage{multirow}
\usepackage[dvipsnames]{xcolor}
\usepackage{bclogo}
\usepackage{comment}
\usepackage{dsfont}
\usepackage{cancel}
\usepackage[normalem]{ulem} %sout
\makeatletter
\def\@fpheader{\relax}
\makeatother

\numberwithin{equation}{subsection}

\newcommand{\Op}{\mathcal{O}}
\newcommand{\isl}{\mathfrak{isl}}
\newcommand{\gca}{\mathfrak{gca}}
\newcommand{\bms}{\mathfrak{bms}}
\newcommand{\vir}{\mathfrak{vir}}
\newcommand{\e}[1]{\operatorname{e}^{#1}}

\renewcommand{\d}{\operatorname{d}\!}
\newcommand{\at}[3]{\,\rule[-#1pt]{0.4pt}{#2em}\raisebox{-#1pt}{\tiny\,$#3$}}

\begin{document}

\title{\vspace*{2cm} Semi-classical BMS-blocks from the Oscillator Construction}

\author{Martin Ammon,}
\author{Se\'an Gray,}
\author{Claire Moran,}
\author{Michel Pannier,}
\author{Katharina W\"olfl}
\affiliation{Theoretisch-Physikalisches Institut, Friedrich-Schiller-Universit\"at Jena,\\
Max-Wien-Platz 1, D-07743 Jena, Germany}

\emailAdd{martin.ammon@uni-jena.de}
\emailAdd{sean.gray@uni-jena.de}
\emailAdd{claire.moran@uni-jena.de}
\emailAdd{michel.pannier@uni-jena.de}
\emailAdd{katharina.woelfl@uni-jena.de}

\vspace{1cm}

\abstract{Flat-space holography requires a thorough understanding of BMS symmetry. We introduce an oscillator construction of the highest-weight representation of the $\mathfrak{bms}_3$ algebra and show that it is consistent with known results concerning the $\mathfrak{bms}_3$ module. We take advantage of this framework to prove that $\mathfrak{bms}_3$-blocks exponentiate in the semi-classical limit, where one of the central charges is large. Within this context, we compute perturbatively heavy, and heavy-light vacuum  $\mathfrak{bms}_3$-blocks.}

\setcounter{tocdepth}{2} 
\maketitle

%%%%%%%%%%%%%%%%%%%%%%%%%%%%%%%%%%%%%%%%%%%%%%%%%%%%%%%%%%%
%%%%%%%%%%%%%%%%%%%%%%%%%%%%%%%%%%%%%%%%%%%%%%%%%%%%%%%%%%%
%%%%%%%%%%%%%%%%%%%%%%%%%%%%%%%%%%%%%%%%%%%%%%%%%%%%%%%%%%%

\section{Introduction}
The Holographic Principle promises a great amount of insight into a possible formulation of quantum gravity \cite{Stephens:1993an,Susskind:1994vu,Bousso:2002ju}. Over the past two decades, a number of statements concerning quantum gravity could be made rigorous within the framework of the AdS/CFT correspondence \cite{Maldacena:1997re,Witten:1998qj,Gubser:1998bc}. Low-dimensional examples of the duality, such as the AdS$_3$/CFT$_2$ correspondence, have proven to be valuable toy models; explicit computations are possible on both sides of the duality and can be matched, see \cite{Eberhardt:2019ywk} for a recent example.

In AdS$_3$/CFT$_2$ the central charge of the two-dimensional boundary CFT is related to the inverse of Newton's constant in the bulk theory \cite{Brown:1986nw}. Remarkably, semi-classical quantum-gravitational phenomena generically emerge from universal properties of two-dimensional conformal field theories at large central charge; in addition to the Virasoro symmetry of the field theory, this statement requires a few reasonable assumptions, including modular invariance, locality, and a sparse spectrum of operators whose conformal dimension remain bounded for large central charge. At finite temperature such CFTs display universal thermodynamics \cite{Hartman:2014oaa} which is in agreement with computations in thermal AdS or hairless BTZ black holes \cite{Kraus:2006wn}. Similar universal behaviour holds for one- and two-point correlation functions of the bounded operators, up to exponentially suppressed correction terms \cite{Kraus:2017kyl}. So-called heavy-light conformal blocks show the approximate thermal nature of heavy operators, which may be viewed as black-hole microstates of the dual gravity side \cite{Fitzpatrick:2015zha}; these blocks may be computed with Wilson-line networks on the dual gravity side \cite{Bhatta:2016hpz,Besken:2016ooo}. 

However, any conclusion made in AdS$_3$/CFT$_2$ is naturally followed by the question to which extent the insight can be carried over to more realistic cases, such as higher-dimensional models or dualities involving gravity in asymptotically flat spacetimes -- the scope of the present work is in connection to the latter aspect. 

Flat-space holography is the moniker given to the expected duality between quantum gravity in asymptotically flat spacetimes and BMS-invariant field theories \cite{Barnich:2010eb,Bagchi:2010zz}. It remains a challenging task to identify an explicit example of such a correspondence; for recent results regarding flat-space holography see for example \cite{Bagchi:2010zz,Duval:2014uva,Barnich:2012xq,Barnich:2012rz,Bagchi:2012xr,Bagchi:2013lma,Bagchi:2014iea,Fareghbal:2013ifa,Detournay:2014fva,Barnich:2015mui,Barnich:2015sca,Bagchi:2015wna,Hosseini:2015uba,Basu:2015evh,Campoleoni:2016vsh,Bagchi:2016bcd,Lodato:2016alv,Oblak:2016eij,Afshar:2016kjj,Oblak:2017ect,Banerjee:2017gzj,Grumiller:2017sjh,Basu:2017aqn,Jiang:2017ecm,Ammon:2017vwt,Hijano:2019qmi,Merbis:2019wgk,Godet:2019wje, Geiller:2020edh, Geiller:2020okp}.

A key ingredient in flat-space holography is the infinite-dimensional Lie algebra $\bms_3$, which is the asymptotic symmetry algebra of three-dimensional gravity with vanishing cosmological constant,  at null infinity \cite{Bondi:1962px,Sachs:1962zza}. The $\bms_3$ algebra is generated by super-rotations $L_n$ and super-translations $M_n$, with $n\in\mathbb{Z}$, and contains two central charges $c_{\text{\tiny L}},c_{\text{\tiny M}}\in\mathbb{R}$  \cite{Barnich:2006av}. Classical pure Einstein gravity demands $c_{\text{\tiny L}}=0$; a non-zero $c_{\text{\tiny L}}$  may be present in parity-violating gravitational theories, e.g. in pure Einstein gravity with an additional gravitational Chern-Simons term \cite{Bagchi:2012yk,Bagchi:2018ryy}.

Assuming the existence of a flat-space holographic duality, the large central charge limit $c_\text{\tiny M}\rightarrow\infty$ of a BMS-invariant field theory should correspond to a semi-classical theory of gravity in asymptotically flat spacetime. Some results regarding BMS-invariant theories concern the bootstrap program \cite{Bagchi:2016geg,Bagchi:2017cpu,Chen:2020vvn}, $\bms_3$ characters and modular invariance \cite{Bagchi:2019unf,Oblak:2015sea,Garbarz:2015lua}, torus $\bms_3$-blocks \cite{Bagchi:2020rwb}, as well as the computation of heavy-light $\bms_3$-blocks using the monodromy method \cite{Hijano:2018nhq}. 

In the context of flat-space holography, heavy-light $\bms_3$-blocks contain information about probe fields in non-trivial asymptotically flat spacetimes in a dual theory of Einstein gravity. The cosmological solutions generated by heavy operators are then quotients of Minkowski spacetimes and are labelled by their mass $M$ given by $8\mathrm{G}_{\text{\tiny N}}M=24\xi /c_{\text{\tiny M}}-1$, and angular momentum $J=\Delta+c_\text{\tiny L}/24$, where $\Delta$ and $\xi$ denote the scaling dimension and rapidity of the heavy operator, respectively \cite{Hijano:2018nhq}, and $\mathrm{G}_{\text{\tiny N}}$ is Newton's constant, which is related to the central charge by $c_{\text{\tiny M}} = 3 /\mathrm{G}_\text{\tiny N}$.

The $\bms_3$ algebra is related to two copies of the Virasoro algebra by suitable \.{I}nönü-Wigner contractions: there exist two such contractions, known as the \emph{ultra-relativistic} and \emph{non-relativistic} limits; both contractions lead to isomorphic symmetry algebras but with different representation theories \cite{Bagchi:2010zz}. One typically finds unitary (induced) representations by ultra-relativistic contraction of the conformal symmetry algebra \cite{Campoleoni:2016vsh}.

The oscillator construction may be employed in order to express highest-weight representations of the Virasoro algebra in terms of oscillator variables \cite{Gervais1985OscillatorRO,Zamolodchikov:1986}; we review this construction in section \ref{sec:rev_virasoro}. This formulation provides an alternative approach to the calculation of various quantities of two-dimensional CFTs, such as correlation functions and conformal blocks, and has recently been utilised in the context of the eigenstate thermalisation hypothesis \cite{Besken:2019bsu} and the proof of Virasoro-block exponentiation \cite{Besken:2019jyw}. 

In the present work we introduce an oscillator construction for highest-weight representations of $\bms_3$.\footnote{Such representations of $\bms_3$ are related to non-relativistic \.{I}nönü-Wigner contractions of highest-weight representations of the conformal algebra; the symmetry algebra under these circumstances is commonly referred to as the Galilean conformal algebra, $\gca_2$ for short. Such representations have been used in the context of non-relativistic Newton-Cartan gravity \cite{Bagchi:2009my}. In contrast, the induced representations mentioned above may be suited for ultra-relativistic Carrollian gravity.}
In section \ref{sec:OscConstr_bms} we express the $\bms_3$ generators in terms of oscillator variables and define a suitable measure on the function space. The origin of our expressions is a non-relativistic limit of a linear-dilaton like theory. The oscillator formalism allows us to prove the exponentiation of $\bms_3$-blocks in the semi-classical limit, which we do in section \ref{sec:SemiClassBlocks}. Moreover, in section \ref{sec:SemiClassBlocks} we also demonstrate the applicability of the oscillator construction by calculating a perturbatively heavy vacuum $\bms_3$-block as well as a heavy-light vacuum $\bms_3$-block.

%%%%%%%%%%%%%%%%%%%%%%%%%%%%%%%%%%%%%%%%%%%%%%%%%%%%%%%%%%%
%%%%%%%%%%%%%%%%%%%%%%%%%%%%%%%%%%%%%%%%%%%%%%%%%%%%%%%%%%%
%%%%%%%%%%%%%%%%%%%%%%%%%%%%%%%%%%%%%%%%%%%%%%%%%%%%%%%%%%%

\section{Oscillator Construction of the Virasoro Algebra}\label{sec:rev_virasoro}
Before approaching the $\bms_3$ oscillator construction it is advantageous to first introduce the oscillator construction and its meaning for two-dimensional conformal field theories. The conformal symmetry algebra decomposes into a direct sum of two copies of the Virasoro algebra: one holomorphic and one anti-holomorphic copy. In the following review we will focus on the holomorphic sector but the analysis straightforwardly extends to the anti-holomorphic sector. 

\subsection{Highest-Weight Representation of the Virasoro Algebra}
The generators of the Virasoro algebra, which we denote $L^\mathfrak{vir}_m,$ obey the commutation relations 
\begin{equation}\label{eq:virasoroalgebra}
    [L^\vir_m, L^{\vir}_n] = (m-n)L^\mathfrak{vir}_{m+n} + \frac{c}{12}m\del{m^2-1}\delta_{m,-n}\,,
\end{equation}
where $c$ denotes the central charge and $m,n\in\mathbb{Z}$. We consider highest-weight representations of the above Virasoro algebra; such representations are built from a primary state $\ket{h}$ with conformal weight $h$, satisfying $L^{\vir}_0 \ket{h} = h\ket{h}$ and $L^{\vir}_m \ket{h} = 0$ for $m\geq1$. 

Descendant states are generated by basis vectors of the form
\begin{equation}\label{eq:basisdesc}
     | \del{ m_1, \hdots, m_k }; h \rangle = \left(\prod\limits_{i=1}^k L^{\vir}_{-m_i}\right) |h\rangle
\end{equation}
with $m_1 \geq \hdots\geq m_k\geq1$. A primary state and its descendants span a vector space called the Verma module $  \mathscr{V}_{c,h}$. Imposing that the central charge $c$ is a real number, as well as the adjoint relation $\big(L^{\vir}_m\big)^\dagger = L^{\vir}_{-m}$, there exists a unique Hermitian product $\langle q|p \rangle$ on the Verma module $\mathscr{V}_{c,h}$ for which a primary state has unit norm. For $c,h>0$ and in the absence of singular states the Hermitian product is positive definite and the corresponding representation is irreducible and unitary. 

\subsection{Oscillator Construction}
We now employ the oscillator construction to study the highest-weight representation of the Virasoro algebra. In this context we write the Virasoro generators in cursive lowercase and they take the form\footnote{These generators are obtained by considering a linear-dilaton conformal field theory. See appendix A of \cite{Besken:2019bsu} for further details.}
\begin{subequations}
\begin{align}
    \ell_0 &= h + \sum_{n=1}^\infty n u_n \partial_{u_n}\,,\label{eq:l0}\\
    \ell_k &= \sum_{n=1}^\infty n u_n \partial_{u_{n+k}} - \frac14 \sum_{n=1}^{k-1}\partial_{u_n} \partial_{u_{k-n}} + \del{\mu k + i \lambda} \partial_{u_k}\,, \label{eq:lk}\\
    \ell_{-k} &= \sum_{n=1}^\infty (n+k)u_{n+k} \partial_{u_n} - \sum_{n=1}^{k-1} n(k-n)u_n u_{k-n} + 2k\del{\mu k - i \lambda}u_k\,,\label{eq:l-k}
\end{align}
\end{subequations}
where $k\ge 1$ and the constants $\lambda,\mu\in \mathbb{R}$ are related to the central charge and conformal weight by\footnote{To gain access to Verma modules with $h< (c-1)/24$ the constant $\lambda$ must be analytically continued to imaginary values. Similarly, minimal models with $c<1$ require an analytic continuation in $\mu$.}
\begin{align}
    c=1+24\mu^2\, \qquad \mathrm{and} \qquad h=\lambda^2+\mu^2\,.
\end{align}
In this language, the states of a Verma module $\mathscr{V}_{c,h}$ are mapped to functions which depend on a full set of infinitely many representation space variables $u_n\in\mathbb{C}$ with $n\in\mathbb{N}$; in accordance with previous literature \cite{Gervais1985OscillatorRO,Zamolodchikov:1986,Besken:2019bsu,Besken:2019jyw} we will refer to these variables as oscillator variables. Equivalence between the oscillator construction and the formalism presented in the previous subsection follows from the definition
\begin{equation}
    f_p(u) \equiv \braket{u|p},
\end{equation}
where $\bra{u}\equiv \ket{\bar{u}}^\dagger$ is a generalised coherent state of the Verma module $\mathscr{V}_{c,h}$; $\ket{p}$ is a generic state of the Verma module, and we omit the index of the oscillator variables to indicate a full set. The presence of a derivative in $\ell_0$ makes it natural that the primary state of $\mathscr{V}_{c,h}$ maps to a constant, which we choose to be unity, i.e. 
\begin{equation}
    f_h(u) = \braket{u|h} \equiv \mathbf{1}\,.
\end{equation}
The basis vectors $|\!\del{ m_1, \dots, m_k }; h \rangle$, which were defined in equation \eqref{eq:basisdesc}, are mapped to 
\begin{equation}\label{basisVirasoro}
    \langle u| \del{\prod\limits_{i=1}^k L^{\vir}_{-m_i} }|h\rangle = \left(\prod\limits_{i=1}^k \ell_{-m_i}\right) \cdot \mathbf{1}\,.
\end{equation}
Given the form of $\ell_{-k}$ in \eqref{eq:l-k}, the basis vector on the right-hand side of the above equation is a polynomial in the oscillator variables.

The dual function is defined by $\overline{f_q(u)} = \overline{\braket{u|q}} \equiv \braket{q|\bar{u}}$, where  $\ket{q}$ is a state in $\mathscr{V}_{c,h}$ and the overline operation acts as $u_n\mapsto\bar{u}_n$ as well as complex conjugation. Such dual functions are acted on by $\bar{\ell}_{-m}$, which results in the action $\langle q|L^{\vir}_n|\bar{u}\rangle$. Note that the barred quantities belong to the same holomorphic sector of the two-dimensional conformal algebra as the non-barred quantities; in other words, the bar on oscillator-dependent quantities should not be confused with the bar which commonly denotes the anti-holomorphic sector of the conformal algebra.

Unitarity requires the adjoint property  $\ell^\dagger_m = \ell_{-m}$, which is defined by the Hermitian product; the appropriate expression in the oscillator construction may be found by inserting a completeness relation
\begin{equation}\label{eq:completesetVirasoro}
    \int_{\mathbb{C}^\infty}[ \dd^2 u ]_h \ket{\bar{u}}\bra{u} = \mathds{1}
\end{equation}
into the product $\braket{q|p}$ defined in the previous subsection. The Hermitian product of the Virasoro oscillator construction thus reads
\begin{align}\label{hermitianprodVirasoro}
    \del{f_q, f_p } = \int_{\mathbb{C}^\infty} [\dd^2 u]_h \, \overline{f_q(u)} f_p(u)\,,
\end{align}
with the measure given by
\begin{align}\label{virasoromeasure}
    [\operatorname{d}^2\!u]_h = \prod_{n=1}^\infty \dd^2 u_n \frac{2n}{\pi} e^{-2n u_n \bar{u}_n}\,,
\end{align}
where $\dd^2 u_n = \dd u_n \dd \bar{u}_n$. Using the Hermitian product given in \eqref{hermitianprodVirasoro}, monomials of oscillator variables form an orthogonal basis which satisfies  
\begin{equation}
       \del{ u_1^{m_1} u_2^{m_2}\cdots, u_1^{\tilde{m}_1} u_2^{\tilde{m}_2}\cdots  } = \prod_{n=1}^\infty \frac{m_n!}{(2n)^{m_n}} \delta_{m_n,\tilde{m}_n} \,.
\end{equation}
The above orthogonality relation proves to be useful when evaluating certain quantities, for instance the Gram matrix \cite{Besken:2019bsu}.

%%%%%%%%%%%%%%%%%%%%%%%%%%%%%%%%%%
%%%%%%%%%%%%%%%%%%%%%%%%%%%%%%%%%%%%%%%%%%%%
%%%%%%%%%%%%%%%%%%%%%%%%%%%%%%%%%%

\subsection{Correlators and Wave Functions}
Building upon the Virasoro symmetry and its representations, a two-dimensional conformal field theory is defined in terms of its operators and correlation functions. A holomorphic primary operator $\mathcal{O}_h(z)$ is defined via the operator-state correspondence as
\begin{equation}
    \ket{h} \equiv \lim_{z\to0}\mathcal{O}_h(z)\ket{0}\,,
\end{equation}
where $z$ is a holomorphic coordinate on the complex plane and the vacuum state $\ket{0}$ is characterised by $h=0$ and $L^\vir_{m} \ket{0}=0$ for $m\geq-1$. Virasoro generators $L^\vir_m$ act on $\Op_h(z)$ via
\begin{equation}\label{eq:actionLn}
    [L^{\vir}_m,\mathcal{O}_{h}(z)] = -\mathcal{L}^{\vir}_m\, \mathcal{O}_{h}(z)\,,
\end{equation}
where $\mathcal{L}^{\vir}_m$ is a differential operator given by 
\begin{equation}
\mathcal{L}^{\vir}_m=-z^{m+1}\partial_{z}-(m+1)h z^m\,.
\end{equation}
Note that $\mathcal{L}^{\vir}_m$ satisfies the Virasoro algebra without a central extension. 

%%%%%%%%%%%%%%%%%%%%%%%%%%%%%%%%%%%%%%

\subsubsection{Two-point Correlation Functions}
The two-point correlation function is given by $\braket{0| \Op_{h_1}(z_1) \Op_{h_2}(z_2) | 0}$; inserting a complete set of states as defined in equation \eqref{eq:completesetVirasoro} results in the expression
\begin{equation}
    \braket{0| \Op_{h_1}(z_1) \Op_{h_2}(z_2) | 0} = \int_{\mathbb{C}^\infty} [\dd^2u]_h \, \braket{0 | \Op_{h_1}(z_1) | \bar{u}}\braket{u | \Op_{h_2}(z_2) | 0}\,.
\end{equation}
The above result defines the level-one wave functions 
\begin{subequations}
    \begin{align}
        \psi_{h_2;h}(z_2;u) &= \langle u | \mathcal{O}_{h_2}(z_2) |0 \rangle \, , \\
        \chi_{h_1;h}(z_1;\bar{u}) &= \langle 0 | \mathcal{O}_{h_1}(z_1) |\bar{u} \rangle\,,
    \end{align}
\end{subequations}
where the subscripts $h_i$ denote the conformal dimensions of the external operators $\mathcal{O}_{h_i}$ while $h$ labels the Verma module $\mathscr{V}_{c,h}$. When treated individually we refer to $\psi_{h_2;h}(z_2;u)$ as the wave function and $\chi_{h_1;h}(z_1;\bar{u})$ as the dual wave function; the relationship between them is
\begin{equation}
    \chi_{h_1;h}(z_1;\bar{u}) = z_1^{-2h_1} \, \overline{\psi_{h_1;h}(z_1^{-1};u)}\,,
\end{equation}
which follows from the property that bra- and ket-states of a conformal field theory are related by a coordinate inversion on the complex plane. Since the barred oscillator-dependent quantities belong to the same holomorphic sector of the conformal algebra as the non-barred quantities, the dual wave function is a function of the holomorphic coordinates $z_i$. 

As a consequence of $L^{\vir}_{m}\ket{0}=0$ for $m\geq -1$, together with \eqref{eq:actionLn}, the level-one wave functions satisfy the differential equations
\begin{subequations}\label{lvl1eqsVira}
    \begin{align}
    \del{\ell_m^{(h)} + \mathcal{L}^{(h_2)}_m } \psi_{h_2;h}(z_2;u) &=0 \,,  \label{wave eq psi-lvl1 Virasoro}\\
    \del{\bar{\ell}^{(h)}_m - {\mathcal{L}}_{-m}^{(h_1)} }\chi_{h_1; h}(z_1;\bar{u})&=0\, .\label{wave eq chi-lvl1 Virasoro}
\end{align}
\end{subequations}
Since the level-one wave functions connect oscillator variables and coordinates of the complex plane, we here introduce the superscripts $h_i$ and $h$ to emphasise the generators' respective domains of action. Note that the above differential equations allow for non-trivial solutions only if $h_1=h $ or $h_2=h$. 
\subsubsection{Three-point Correlation Functions and Conformal Blocks}
Inserting a completeness relation into the definition of a three-point correlation function, in two different ways, gives
\begin{subequations}
\begin{align}
    \langle 0|\mathcal{O}_{h_1}(z_1) \mathcal{O}_{h_2}(z_2) &\mathcal{O}_{h_3}(z_3) |0\rangle \notag\\
    &= \int_{\mathbb{C}^\infty} [\dd^2 u]_h \, \braket{0|\mathcal{O}_{h_1}(z_1)|\bar{u}}\braket{u| \mathcal{O}_{h_2}(z_2) \mathcal{O}_{h_3}(z_3) |0}\\
    &= \int_{\mathbb{C}^\infty} [\dd^2 u]_h \, \braket{0|\mathcal{O}_{h_1}(z_1) \mathcal{O}_{h_2}(z_2)|\bar{u}}\braket{u| \mathcal{O}_{h_3}(z_3) |0}\,.
\end{align}
\end{subequations}
The structure of the above expressions may be used to define the level-two wave functions
\begin{subequations}
    \begin{align}
      \psi_{h_1, h_2;h}(z_1,z_2;u) &= \langle u | \mathcal{O}_{h_1}(z_1) \mathcal{O}_{h_2}(z_2) |0 \rangle \, , \label{wave eq psi-lvl2 Virasoro} \\
      \chi_{h_3,h_4;h}(z_3,z_4;\bar{u}) &= \langle 0 | \mathcal{O}_{h_4}(z_4) \mathcal{O}_{h_3}(z_3) |\bar{u} \rangle \, \label{wave eq chi-lvl2 Virasoro} .
\end{align}
\end{subequations}
Following the same reasoning as in the previous paragraph the level-two wave functions are related by 
\begin{equation}
     \chi_{h_3, h_4;h}(z_3, z_4; \bar{u}) = z_3^{-2h_3} \, z_4^{-2h_4} \, \overline{\psi_{h_3, h_4;h}(z_3^{-1}, z_4^{-1};u)} \, , 
\end{equation}
and they satisfy the differential equations
\begin{subequations}\label{lvl2eqsVira}
    \begin{align}
    \del{\ell^{(h)}_m + \mathcal{L}^{(h_1)}_m + \mathcal{L}^{(h_2)}_m } \psi_{h_1, h_2;h}(z_1,z_2;u) &=0\,,\\
    \del{\bar{\ell}^{(h)}_m - {\mathcal{L}}_{-m}^{(h_3)}  - {\mathcal{L}}_{-m}^{(h_4)} }\chi_{h_3, h_4;h}(z_3,z_4;\bar{u})&=0\, .
\end{align}
\end{subequations}

We may now appropriately express conformal blocks in terms of wave functions. Virasoro blocks are defined by the four-point function $\braket{ 0| \mathcal{O}_{h_4}(z_4) \mathcal{O}_{h_3}(z_3) \mathcal{P}_{h} \mathcal{O}_{h_1}(z_1) \mathcal{O}_{h_2}(z_2) |0 }$, where $\mathcal{P}_h$ denotes the projector onto the Verma module $\mathscr{V}_{c,h}$. It follows directly from the argumentation and definitions presented in this subsection that a conformal block reads 
\begin{equation}
\begin{split}
    \langle 0| \mathcal{O}_{h_4}(z_4) \mathcal{O}_{h_3}(z_3) &\mathcal{P}_{h} \mathcal{O}_{h_1}(z_1) \mathcal{O}_{h_2}(z_2) |0 \rangle \\ 
    &= \int_{\mathbb{C}^\infty} [\dd^2u]_h\, \chi_{h_3,h_4;h}(z_3,z_4;\bar{u}) \, \psi_{h_1, h_2;h}(z_1,z_2;u)\,.
   \end{split}
\end{equation}
The operator-state correspondence allows for the primary and descendant states of the Verma module $\mathscr{V}_{c,h}$ to be interpreted as internal operators which arise due to the operator product expansion of the four-point correlation function.

Although the differential equations \eqref{lvl1eqsVira} allow for solutions for general values of $h$ and $c$, the level-two equations \eqref{lvl2eqsVira} have known closed-form solutions for the case $h_i = 1/16$ and $c=1$, only \cite{Zamolodchikov:1986}. Nevertheless, in the semi-classical limit, i.e. $c\to\infty$, it is possible to find approximate solutions for the level-two wave functions, and by extension the Virasoro conformal blocks \cite{Besken:2019jyw}.

%%%%%%%%%%%%%%%%%%%%%%%%%%%%%%%%%%%%%%%%%%%%%%%%%%%%%%%%%%%
%%%%%%%%%%%%%%%%%%%%%%%%%%%%%%%%%%%%%%%%%%%%%%%%%%%%%%%%%%%
%%%%%%%%%%%%%%%%%%%%%%%%%%%%%%%%%%%%%%%%%%%%%%%%%%%%%%%%%%%

\section{Oscillator Construction of the Highest-Weight Representation of \texorpdfstring{$\bms_3$}{bms3}}\label{sec:OscConstr_bms}
The $\bms_3$ algebra is a semi-direct sum of one Virasoro algebra and an infinite-dimensional Abelian algebra of super-translations. Hence, in contrast to the Virasoro case, the $\bms_3$ algebra cannot be decomposed into commuting sectors; thus it needs to be treated \emph{en bloc}. In this section we first review highest-weight $\bms_3$ modules, after which we present our oscillator construction of such modules. We also discuss the computation of correlation functions and $\bms_3$-blocks in the language of oscillator variables and wave functions. 

%%%%%%%%%%%%%%%%%%%%%%%%%%%%%%%%
%%%%%%%%%%%%%%%%%%%%%%%%%%%%%%%%%

\subsection{Modules of the \texorpdfstring{$\bms_3$}{bms3} Algebra}
The $\bms_3$ algebra is generated by $L_n$ and $M_n$, and defined by the Lie bracket
\begin{subequations}\label{eq:bms3}
\begin{align}
    [L_m,L_n] &= (m-n)L_{m+n} + \frac{c_{\text{\tiny L}}}{12}m(m^2-1)\delta_{m,-n}\,, \label{bmsLL}\\
    [L_m,M_n] &= (m-n)M_{m+n} + \frac{c_{\text{\tiny M}}}{12}m(m^2-1)\delta_{m,-n}\,, \label{bmsLM}\\
    [M_m,M_n] &= 0 \, , \label{bmsMM}
\end{align}
\end{subequations}
where $c_{\text{\tiny L}}$ and $c_{\text{\tiny M}}$ are central charges, which we take to be non-negative real numbers, and $m,n\in\mathbb{Z}$. In the context of asymptotically flat gravity in three spacetime dimensions $L_n$ are the generators of super-rotations and the $M_n$ generate super-translations.

We define a primary state $\ket{\Delta,\xi}$ to be a state that satisfies the eigenvalue equations
\begin{equation}\label{LMeigenvalue}
    L_0 |\Delta, \xi \rangle = \Delta  |\Delta, \xi \rangle \qquad \text{and} \qquad M_0 |\Delta, \xi \rangle = \xi  |\Delta, \xi \rangle\,,
\end{equation} 
where $\Delta$ is the scaling dimension and $\xi$ is the rapidity, as well as
\begin{equation}\label{LMprimary}
    L_n |\Delta, \xi \rangle = 0 \qquad \text{and} \qquad M_n |\Delta, \xi \rangle = 0 \, ,
\end{equation} 
for $n>0$. Analogously to the construction of a Verma module, a $\bms_3$ module $\mathscr{B}^{c_{\text{\tiny L}},c_{\text{\tiny M}}}_{\Delta,\xi}$ is built by acting with an ordered string, of arbitrary length, of operators $L_{-n}$ and $M_{-n}$ with $n>0$ on a state $|\Delta, \xi \rangle$.
Hence, the vector space associated to the highest-weight representation of $\bms_3$ is spanned by the basis vectors 
\begin{equation}\label{eq:basisvecbms}
  | (m_1,\hdots,m_s), (n_1, \hdots, n_l); \Delta, \xi \rangle =  L_{-m_1} \cdots L_{-m_s} M_{-n_1} \cdots M_{-n_l} |\Delta, \xi \rangle\,,  
\end{equation}
where $m_1 \geq \hdots\geq m_s\geq1$ and  $n_1 \geq \hdots\geq n_l\geq1$. 

The Hermitian product $\braket{q|p}$ for $\ket{p},\ket{q}\in\mathscr{B}^{c_{\text{\tiny L}},c_{\text{\tiny M}}}_{\Delta,\xi}$ is uniquely defined by imposing the adjoint relations $L_{n}^\dagger = L_{-n}$ and $M_{n}^\dagger = M_{-n}$. The Hermitian product of two basis vectors is thus given by
\begin{equation}
\begin{split}\label{eq:BMSGram}
    \Big\langle ( m'_i )_{i=1}^{s'},  ( n'_j )_{j=1}^{l'}; \Delta,\xi  \,&\Big| ( m_{i} )_{i=1}^s, ( n_{j} )_{j=1}^l; \Delta, \xi \Big\rangle \\ &=\Big\langle \Delta,\xi\Big|\prod\limits_{j=l'}^{1} M_{n'_j} \prod\limits_{i=s'}^{1} L_{m'_i}   \prod\limits_{i=1}^s L_{-m_i} \prod\limits_{j=1}^l M_{-n_j} \Big|\Delta, \xi \Big\rangle \, ,
\end{split}
\end{equation}
where we used the short-hand notation $(m_i)_{i=1}^s \equiv (m_1,\dots,m_s)$. For the highest-weight representation of $\bms_3$ the Hermitian product is not generically positive semi-definite and hence the corresponding representation will not necessarily be unitary.\footnote{There exists an exceptional case for $c_{\text{\tiny M}}=0$ and $\xi=0$ where the $\mathfrak{bms}_3$ representation constructed above reduces to a  Virasoro highest-weight representation with central charge $c_{\text{\tiny L}}$ and conformal dimension $h=\Delta$, provided that we take a quotient with respect to the null states $M_{-n} |\Delta,0\rangle$ with $n\in\mathbb{N}$ \cite{Bagchi:2009pe}.} We return to this point at the end of the next subsection.

%%%%%%%%%%%%%%%%%%%%%%%%%%%%%%%%%
%%%%%%%%%%%%%%%%%%%%%%%%%%%%%%%%%

\subsection{Oscillator Construction}\label{subsec:OscConstr}
The expressions for the $\bms_3$ generators in terms of the complex oscillator variables $v^{(1)}_n$ and $v^{(2)}_n$, with $n\in\mathbb{N}$, may be found by taking a non-relativistic limit of a two-dimensional linear-dilaton like conformal field theory; we refer the reader to appendix \ref{app:building_oscill_rep} for details. The resulting generators, which we denote in lowercase, read\footnote{A similar looking relationship between $\mathfrak{u}(1)$ current algebras and $\mathfrak{bms}_3$ generators was found in \cite{Afshar:2016kjj}; see also \cite{Ammon:2017vwt} for a higher-spin generalisation of said relationship. }
\begin{subequations}\label{lbms}
\begin{align}
\label{eq:l0bms}
    l_0 &= \Delta + \sum_{n=1}^\infty n \del{v^{(1)}_n \partial_{v^{(1)}_n} + v^{(2)}_n \partial_{v^{(2)}_n}} \,, \\
\begin{split}
        l_k &= \sum_{n=1}^\infty n\del{ {v^{(1)}_n} \partial_{ v^{(1)}_{k+n}} + v^{(2)}_n \partial_{ v^{(2)}_{k+n}} } 
        -\frac14 \sum_{n=1}^{k-1}\partial_{v^{(1)}_n} \partial_{ v^{(2)}_{k-n}} +A_k \partial_{ v^{(1)}_k } + B_k \partial_{ v^{(2)}_k }\label{lkBMS}\,,
\end{split}\\
\begin{split}
        l_{-k} &= \sum_{n=1}^\infty (k+n)\del{ {v^{(1)}_{k+n} } \partial_{ v^{(1)}_{n}} + v^{(2)}_{k+n} \partial_{ v^{(2)}_{n}} }  -4 \sum_{n=1}^{k-1} n(k-n){v^{(1)}_n} { v^{(2)}_{k-n}} + 4 k \hat{B}_k { v^{(1)}_k } + 4 k \hat{A}_k { v^{(2)}_k }\label{l-kBMS}\,,
\end{split}
\end{align}
\end{subequations}
and,
\begin{subequations}\label{mbms}
\begin{align}
    m_0 &= \xi+ \sum_{n=1}^\infty n v^{(1)}_n \partial_{ v^{(2)}_n } \,, \label{mk0BMS}\\
    m_k &= \sum_{n=1}^\infty n v^{(1)}_n \partial_{ v^{(2)}_{k+n} } - \frac18 \sum_{n=1}^{k-1} \partial_{ v^{(2)}_{k-n} } \partial_{ v^{(2)}_{n} } + A_k\partial_{v^{(2)}_k}\,, \label{mkBMS}\\
    m_{-k} &= \sum_{n=1}^\infty (k+n) v^{(1)}_{k+n} \partial_{ v^{(2)}_{n} } - 2 \sum_{n=1}^{k-1} n(k-n){ v^{(1)}_{k-n} } { v^{(1)}_{n} } + 4 k \hat{A}_k{v^{(1)}_k}\,,\label{m-kBMS}
\end{align}
\end{subequations}
for $k>0$. The above generators satisfy the $\bms_3$ commutation relations \eqref{eq:bms3}. In the above expressions we have made the identifications
\begin{subequations}\label{Ak_Bkfull}
\begin{align}\label{Ak_Bk}
{A}_k&=-\frac{i}{2} \sqrt{2\xi-\frac{c_{\text{\tiny M}}}{12}} - k\sqrt{\frac{c_{\text{\tiny M}}}{48}}\,,     & {B}_k &=i\frac{c_{\text{\tiny L}}-2-24\Delta}{48\sqrt{2\xi-\frac{c_{\text{\tiny M}}}{12}}}-  k\frac{c_{\text{\tiny L}}-2}{48\sqrt{\frac{c_{\text{\tiny M}}}{12}}}\,,\\
\label{hatAk_hatBk}
\hat{A}_k&=\frac{i}{2}\sqrt{2\xi-\frac{c_{\text{\tiny M}}}{12}}- k\sqrt{\frac{c_{\text{\tiny M}}}{48}}\,,     & \hat{B}_k&=-i\frac{c_{\text{\tiny L}}-2-24\Delta}{48\sqrt{2\xi-\frac{c_{\text{\tiny M}}}{12}}} - k\frac{c_{\text{\tiny L}}-2}{48\sqrt{\frac{c_{\text{\tiny M}}}{12}}}\,.
\end{align}
\end{subequations}
If $\xi \geq c_{\text{\tiny M}}/24$, the above coefficients are related by complex conjugation $\hat{A}_k=A^*_k$ and $\hat{B}_k=B_k^*$; in this case the adjoint property $l_n^\dagger=l_{-n}$ holds. However, for $\xi < c_{\text{\tiny M}}/24$ all coefficients are real and hence independent of each other; preserving the adjoint property requires analytic continuation, as discussed in section \ref{sec:analyticBEHAVIOUR}. For simplicity, unless otherwise stated, we will assume $\xi \geq c_{\text{\tiny M}}/24$.

Analogously to the Verma module, a state $\ket{p}\in\mathscr{B}^{c_{\text{\tiny L}},c_{\text{\tiny M}}}_{\Delta,\xi}$ is mapped to a function via
\begin{equation}
    f_p(v) \equiv \braket{v|p}\,,
\end{equation}
where $\bra{v}\equiv\ket{\bar{v}}^\dagger$ is a generalised coherent state of the $\bms_3$ module, and $v$ denotes the collective set of all oscillator variables $v^{(1)}_n$ and $v^{(2)}_n$.\footnote{To flesh out the notation we could equivalently write $\ket{v}=\ket{v^{(1)}, v^{(2)}}$. We drop the index to indicate the infinite collection.} It follows from the form of the generators \eqref{lbms} and \eqref{mbms} that the requirements for a primary state are fulfilled by a constant function, hence we choose 
\begin{equation}\label{BMSprimary}
    f_{\Delta,\xi}(v) = \braket{v|\Delta,\xi} \equiv \mathbf{1}\,.
\end{equation}
 The properties \eqref{LMeigenvalue} and \eqref{LMprimary} translate to
\begin{subequations}
\begin{align}
    l_0 \cdot \mathbf{1}&=\Delta\,,  & m_0 \cdot \mathbf{1}&=\xi\,,\\
    l_k \cdot  \mathbf{1}&=0\,,   & m_k \cdot \mathbf{1}&=0\,,  
\end{align}
\end{subequations}
for $k>0$. The basis vectors of the form \eqref{eq:basisvecbms} are thus given in terms of the polynomials $\prod_{i=1}^s l_{-m_i} \prod_{j=1}^l m_{-n_j} \cdot \mathbf{1}$. Additionally, we define $\overline{f_q(v)} = \overline{\braket{v|q}}\equiv\braket{q|\bar{v}}$ for the state $\ket{q}\in\mathscr{B}^{c_{\text{\tiny L}},c_{\text{\tiny M}}}_{\Delta,\xi}$, where the overline acts as $v^{(i)}_n\mapsto\bar{v}^{(i)}_n$ and complex conjugation. The barred functions are acted on with the generators $\bar{l}_{-n}$ and $\bar{m}_{-n}$, which in terms of states is expressed as $\langle q|L_n |\bar{v}\rangle$ and $\langle q|M_n|\bar{v}\rangle$, respectively.

Finally, the adjoints $l_n^\dagger = l_{-n}$ and $m_n^\dagger = m_{-n}$ specify the unique Hermitian product of the oscillator construction of the highest-weight representation of $\bms_3$ to be
\begin{equation}\label{eq:inner_product}
    (f_q,g_p) = \int_{\mathbb{C}^\infty} [\dd^2 v]_{\Delta,\xi} \,\overline{f_q(v)}g_p(v)\,,
\end{equation}
where the measure is given by
\begin{equation}\label{eq:measure}
    [\dd^2 v ]_{\Delta,\xi}=\prod_{n=1}^{\infty} 16n^2 \exp[-4n\del{v^{(1)}_{n}\bar{v}^{(2)}_{n}+v^{(2)}_{n}\bar{v}^{(1)}_{n}}] \, \dd^{2}v^{(1)}_{n} \, \dd^{2}v^{(2)}_{n}\,,
\end{equation}
with $\dd^2 v^{(i)}_n = \dd v^{(i)}_n\dd \bar{v}^{(i)}_n$. The motivation for the form of the measure $[\dd^2 v]_{\Delta,\xi}$ is deferred to the appendix \ref{app:subsec:operators_measure}. We arrived at the expression \eqref{eq:inner_product} in the same way as one does for Virasoro, i.e. by inserting a complete set of states of the form 
\begin{equation}\label{bms3completeness}
    \int_{\mathbb{C}^\infty} [\dd^2 v]_{\Delta,\xi} \,\ket{\bar{v}}\bra{v} = \mathds{1}   
\end{equation}
into the Hermitian product $\braket{q|p}$. 

The Hermitian product of oscillator monomials is given by the formula
\begin{align}\label{eq:orthogonality}
    \left(\left(v^{(1)}_m\right)^{a}\left(v^{(2)}_m\right)^{b}\,,\left(v^{(1)}_m\right)^{c}\left(v^{(2)}_m\right)^{d}\right)=\frac{a!\,b!}{(4m)^{a+b}}\delta_{a,d}\delta_{b,c}\,,
\end{align}
which is a result of a calculation that is sketched in appendix \ref{app:subsec:operators_measure}. By calculating the Hermitian product of basis vectors we find the Gram matrix; in the oscillator construction a general element of the form \eqref{eq:BMSGram} is given by 
\begin{equation}\label{eq:grammatrixInner}
    \left(l_{-m'_1}\cdots l_{-m'_{s'}} m_{-n'_1}\cdots m_{-n'_{l'}}\cdot\mathbf{1}\,,\,l_{-m_1}\cdots l_{-m_s} m_{-n_1}\cdots m_{-n_l}\cdot\mathbf{1}\right)\,.
\end{equation}
The basis vectors are polynomials in $v^{(1)}_n$ and $v^{(2)}_n$, hence to calculate a specific element of the Gram matrix one may advantageously apply the orthogonality relation \eqref{eq:orthogonality} to their constituent monomials. For the lowest-level $\bms_3$ Gram matrix we obtain the entries
\begin{equation}
\begin{aligned}
\left(l_{-1}\cdot\mathbf{1},l_{-1}\cdot\mathbf{1}\right)&= 2\Delta\,,   & \hspace{1cm} \left(l_{-1}\cdot\mathbf{1},m_{-1}\cdot\mathbf{1}\right)&= 2\xi\,,\\
\left(m_{-1}\cdot\mathbf{1},l_{-1}\cdot\mathbf{1}\right)&=\, 2\xi\,,   & \hspace{1cm} \left(m_{-1}\cdot\mathbf{1},m_{-1}\cdot\mathbf{1}\right)&=0\,.
\end{aligned}
\end{equation}
The above matrix components, as well as the second-level Gram matrix entries (which are not displayed here) match the results of \cite{Bagchi:2019unf}. This serves as a check of our oscillator construction of $\bms_3$. 

The highest-weight representation of $\bms_3$ is in general not unitary. This can be seen from the lowest-level Gram matrix since it has one positive and one negative eigenvalue if $\xi\neq0$; thus the Hermitian product is indefinite and hence the highest-weight representation is non-unitary.

%%%%%%%%%%%%%%%%%%%%%%%%%%%%%%%%%%%%%%%%%%%%%%%%%%%
%%%%%%%%%%%%%%%%%%%%%%%%%%%%%%%%%%%%%%%%%%%%%%%%%%%

\subsection{Correlators and Wave Functions}
We define a $\bms_3$ primary operator $\Op_{\Delta,\xi}(t,x)$ by means of the operator-state correspondence
\begin{equation}
    \ket{\Delta,\xi}\equiv \lim_{t,x\to0}\Op_{\Delta,\xi}(t,x)\ket{0} \, ,
\end{equation}
where $t$ and $x$ are coordinates on the plane. The vacuum state $\ket{0}$ is defined as the primary state with $\Delta=\xi=0$ and $L_{n}\ket{0}=M_{n}\ket{0}=0$ for $n\geq-1$. The generators of $\bms_3$ act on $\Op_{\Delta,\xi}(t,x)$ as
\begin{subequations}\label{bmsdiffopaction}
    \begin{align}
        [L_n,\Op_{\Delta,\xi}(t,x)] &= -\mathcal{L}_n \, \Op_{\Delta,\xi}(t,x)\,, \\
        [M_n,\Op_{\Delta,\xi}(t,x)] &= -\mathcal{M}_n \, \Op_{\Delta,\xi}(t,x)\,,
    \end{align}
\end{subequations}
and the differential operators take the forms \cite{Bagchi:2009ca}
\begin{subequations}\label{curlyLM}
\begin{align}
    \mathcal{L}_n &= -t^{n+1}\partial_t -(n+1) t^n x \partial_x - (n+1) (t^n \Delta + n t^{n-1} x \xi)\,,\label{curlyL}\\
    \mathcal{M}_n &= -t^{n+1}\partial_x -(n+1)\xi t^n \label{curlyM}\,.
\end{align}
\end{subequations}
Note that $\mathcal{L}_n$ and $\mathcal{M}_n$ satisfy the commutation relations \eqref{eq:bms3} with $c_{\text{\tiny L}}=c_{\text{\tiny M}}=0$.

%%%%%%%%%%%%%%%%%%%%%%%%%%%%%%%%%%%%%%%%%%%%%%%%%%%%%

\subsubsection{Two-point Correlation Functions}
Inserting the completeness relation \eqref{bms3completeness} into the definition of a two-point correlation function of primary operators gives 
\begin{equation}\label{2ptBMS}
    \braket{0|\Op_{\Delta_1,\xi_1}(t_1,x_1)\Op_{\Delta_2,\xi_2}(t_2,x_2)|0} = \int_{\mathbb{C}^\infty} [\dd^2 v]_{\Delta,\xi}\, \chi_{\Delta_1,\xi_1;\Delta,\xi}(t_1,x_1;\bar{v}) \psi_{\Delta_2,\xi_2;\Delta,\xi}(t_2,x_2;v)\,,
\end{equation}
where we have defined the level-one wave functions
\begin{subequations}
\begin{align}
    \psi_{\Delta_2,\xi_2;\Delta,\xi}(t_2,x_2;v)&=\langle v|\mathcal{O}_{\Delta_2,\xi_2}(t_2,x_2)|0\rangle\,,\\
    \chi_{\Delta_1,\xi_1;\Delta,\xi}(t_1,x_1;\bar{v})&=\langle 0 | \mathcal{O}_{\Delta_1,\xi_1}(t_1,x_1)|\bar{v}\rangle\,.
\end{align}
\end{subequations}
We will refer to $\psi_{\Delta_2,\xi_2;\Delta,\xi}(t_2,x_2;v)$ as the wave function and $\chi_{\Delta_1,\xi_1;\Delta,\xi}(t_1,x_1;\bar{v})$ as the dual wave function. The subscripts $\Delta_i$ and $\xi_i$ label the scaling dimension and rapidity of the external operators, respectively, while $\Delta$ and $\xi$ label the $\bms_3$ module $\mathscr{B}^{c_{\text{\tiny L}},c_{\text{\tiny M}}}_{\Delta,\xi}$. Using that $L_{n}\ket{0}=M_{n}\ket{0}=0$ for $n\geq-1$, as well as the definitions of the differential operators \eqref{bmsdiffopaction}, we find two sets of differential equations for the wave function $\psi_{\Delta_2,\xi_2;\Delta,\xi}(t_2,x_2;v)$,
\begin{subequations}\label{psilvl1BMS}
\begin{align}
    \left(l^{(\Delta,\xi)}_n + \mathcal{L}^{(\Delta_2,\xi_2)}_n\right)\psi_{\Delta_2,\xi_2;\Delta,\xi}(t_2,x_2;v)&=0\,,\label{psilvl1BMSa}\\
    \left(m^{(\Delta,\xi)}_n+\mathcal{M}^{(\Delta_2,\xi_2)}_n\right)\psi_{\Delta_2,\xi_2;\Delta,\xi}(t_2,x_2;v)&=0\,,\label{psilvl1BMSb}
\end{align}
\end{subequations}
for $n \geq -1$. By similar arguments we find that the dual wave function $\chi_{\Delta_1,\xi_1;\Delta,\xi}(t_1,x_1;\bar{v})$ is constrained by the set of differential equations 
\begin{subequations}\label{chilvl1BMS}
\begin{align}
    \left(\bar{l}^{(\Delta,\xi)}_{n}-\mathcal{L}_{-n}^{(\Delta_1,\xi_1)}\right)\chi_{\Delta_1,\xi_1;\Delta,\xi}(t_1,x_1;\bar{v})&=0\,,\\
    \left(\bar{m}^{(\Delta,\xi)}_{n}-\mathcal{M}_{-n}^{(\Delta_1,\xi_1)}\right)\chi_{\Delta_1,\xi_1;\Delta,\xi}(t_1,x_1;\bar{v})&=0\,,
\end{align}
\end{subequations}
for $n\geq -1$. Above we use superscripts for the generators $l_n$, $m_n$ as well as $\mathcal{L}_n$ and $\mathcal{M}_n$ for the same reason as discussed below equations \eqref{lvl1eqsVira}. The wave functions may be determined by solving both sets of differential equations for $n\in\{-1,0,1,2\}$, which also ensures the validity of the solutions for all $n>2$.\footnote{This can be proven by induction, remembering that for $k>1$ we may express the generators $l_{k+1}$ and $\mathcal{L}_{k+1}$ in terms of commutators $[l_1,l_k]$ and $[\mathcal{L}_1, \mathcal{L}_k]$.}

The differential equations \eqref{psilvl1BMS} and \eqref{chilvl1BMS} have non-trivial solutions only if $\Delta_2 = \Delta$, $\xi_2=\xi$, and $\Delta_1 = \Delta$, $\xi_1 = \xi$, respectively.\footnote{To keep the notation compact we drop degenerate subscripts.} We find the following expressions for the level-one $\bms_3$ wave functions,
\begin{subequations}\label{lvl1_BMSwf}
    \begin{align}
        \psi_{\Delta,\xi}(t_2,x_2;v) &= \exp\left[  4 \hat{A}_1\sum_{n=1}^\infty \del{ t_2^n v^{(2)}_n + n\,x_2\,t_2^{n-1} v^{(1)}_n} + 4\hat{B}_1  \sum_{n=1}^\infty t_2^{n} v^{(1)}_n\right]\,, \\
        \chi_{\Delta,\xi}(t_1,x_1;\bar{v}) &= t_1^{-2\Delta}\e{-2\xi \frac{x_1}{t_1}} \exp\left[ 4 A_1\sum_{n=1}^\infty \del{ t_1^{-n} \bar{v}^{(2)}_n - n\,x_1\,t_1^{-n-1} \bar{v}^{(1)}_n} + 4 B_1 \sum_{n=1}^\infty t_1^{-n} \bar{v}^{(1)}_n\right]\,,
    \end{align}
\end{subequations}
where the coefficients $A_1$, $B_1$ and $\hat{A}_1$, $\hat{B}_1$ are given by equations \eqref{Ak_Bk} and \eqref{hatAk_hatBk} for $k=1$, respectively.\footnote{We detail the solution procedure for $\psi_{\Delta,\xi}(t_2,x_2;v)$ in appendix \ref{sec:lvl1soln}. Note that the level-one wave functions are completely fixed by the differential equations \eqref{psilvl1BMS} and \eqref{chilvl1BMS} for $n \in \{-1,0,1\}$. This is connected to the fact that two-point functions are completely determined by the globally well-defined generators $L_n$ and $M_n$ with $n \in \{-1,0,1\}.$ Nevertheless, the set of $n=2$ differential equations must still be satisfied in order to guarantee a general solution for all $n$.} The relationship between the above level-one wave functions is thus
\begin{align}
    \chi_{\Delta,\xi}(t_1,x_1;\bar{v})=t_1^{-2\Delta} \e{-2\xi \frac{x_1}{t_1}}\overline{\psi_{\Delta,\xi}\left(t_1^{-1},-x_1 t_1^{-2};v\right)}\,.
\end{align}
The above relation can be determined from the expressions \eqref{lvl1_BMSwf} or motivated by the operator-state correspondence for bra-states
\begin{equation}\label{eq:BraKetRelation}
    \bra{\Delta,\xi} = \lim_{t\to\infty} t^{2\Delta} \e{2\xi \frac{x}{t}}\bra{0}\Op_{\Delta,\xi}(t,x)\,.
\end{equation}
Inserting the wave functions \eqref{lvl1_BMSwf} in \eqref{2ptBMS}, power expanding, and using the orthogonality relation \eqref{eq:orthogonality}, we arrive at the expression for the $\bms_3$ two-point correlation function 
\begin{equation}
    \bra{0}\mathcal{O}_{\Delta_1,\xi_1}(t_1,x_1)\mathcal{O}_{\Delta_2,\xi_2}(t_2,x_2)\ket{0}=(t_{1}-t_2)^{-2\Delta}e^{-\frac{2\xi (x_1-x_2)}{(t_1-t_2)}}\,,
\end{equation}
if $\Delta_1=\Delta_2=\Delta$ and $\xi_1=\xi_2=\xi$, and zero otherwise. The above expression is in agreement with previously known results \cite{Bagchi:2009ca}.

\subsubsection{Three-point Correlation Functions and $\bms_3$-blocks}
In terms of our oscillator construction for $\bms_3$ modules, the three-point correlation function reads 
\begin{subequations}\label{3ptbmsWF}
\begin{align}
    \langle0|\Op_{\Delta_1,\xi_1}(t_1,x_1)&\Op_{\Delta_2,\xi_2}(t_2,x_2)\Op_{\Delta_3,\xi_3}(t_3,x_3)|0\rangle\notag\\
    &= \int_{\mathbb{C}^\infty} [\dd^2 v]_{\Delta,\xi} \, \chi_{\Delta_{1,2},\xi_{1,2};\Delta,\xi}(t_{1,2},x_{1,2};\bar{v}) \psi_{\Delta_3,\xi_3;\Delta,\xi}(t_3,x_3;v)\, ,\\
    &= \int_{\mathbb{C}^\infty} [\dd^2 v]_{\Delta,\xi} \, \chi_{\Delta_{1},\xi_{1};\Delta,\xi}(t_{1},x_{1};\bar{v}) \psi_{\Delta_{2,3},\xi_{2,3};\Delta,\xi}(t_{2,3},x_{2,3};v)\,,
\end{align}
\end{subequations}
where we have defined the level-two wave functions by
\begin{subequations}
\begin{align}
    \psi_{\Delta_{2,3},\xi_{2,3};\Delta,\xi}(t_2,x_2, t_3,x_3;v) &= \braket{v|\Op_{\Delta_2,\xi_2}(t_2,x_2)\Op_{\Delta_3,\xi_3}(t_3,x_3)|0}\,, \\
    \chi_{\Delta_{1,2},\xi_{1,2};\Delta,\xi}(t_1,x_{1}, t_2,x_2;\bar{v})&=\braket{0|\Op_{\Delta_2,\xi_2}(t_2,x_2)\Op_{\Delta_1,\xi_1}(t_1,x_1)|\bar{v}}\,,
\end{align}
\end{subequations}
where the shorthand notation $\Delta_{i,j}$ and $\xi_{i,j}$ indicates the dependence on both $\Delta_i$, $\Delta_j$ and $\xi_i$, $\xi_j$. The left- and right-hand sides of \eqref{3ptbmsWF} are related by an insertion of a complete set of states \eqref{bms3completeness}. 

The transformation between bra- and ket-states (\ref{eq:BraKetRelation}) means that the level-two wave functions are related by
\begin{equation}\label{lvl2wftransBMS}
\begin{split}
    &\chi_{\Delta_{3,4},\xi_{3,4};\Delta,\xi}(t_{3},x_{3},t_4,x_4;\bar{v}) \\
    &\quad\quad\quad = t_3^{-2\Delta_3} \e{-2\xi_3 \frac{x_3}{t_3}} t_4^{-2\Delta_4} \e{-2\xi_4 \frac{x_4}{t_4}}\overline{ \psi_{\Delta_{3,4},\xi_{3,4};\Delta,\xi}(t_3^{-1},-x_3 t_3^{-2},t_4^{-1},-x_4 t_4^{-2};v) }\,.
\end{split}
\end{equation}
Following the same reasoning as in the previous subsection we find that the wave function must satisfy the set of differential equations 
\begin{subequations}\label{diffeqpsilvl2BMS}
    \begin{align}
        \del{ l^{(\Delta,\xi)}_n + \mathcal{L}^{(\Delta_1,\xi_1)}_n + \mathcal{L}^{(\Delta_2,\xi_2)}_n} \psi_{\Delta_{1,2},\xi_{1,2};\Delta,\xi}(t_{1},x_{1},t_2,x_2;v) &=0\,, \label{psiBMSlvl2-lequation}\\
        \del{ m^{(\Delta,\xi)}_n + \mathcal{M}^{(\Delta_1,\xi_1)}_n + \mathcal{M}^{(\Delta_2,\xi_2)}_n} \psi_{\Delta_{1,2},\xi_{1,2};\Delta,\xi}(t_{1},x_{1},t_2,x_2;v) &=0\,, \label{psiBMSlvl2-mequation}
    \end{align}
\end{subequations}
and similarly the dual wave function must obey
\begin{subequations}
    \begin{align}
        \del{ \bar{l}^{(\Delta,\xi)}_n - \mathcal{L}^{(\Delta_3,\xi_3)}_{-n} - \mathcal{L}^{(\Delta_4,\xi_4)}_{-n}} \chi_{\Delta_{3,4},\xi_{3,4};\Delta,\xi}(t_{3},x_{3},t_4,x_4;\bar{v}) &=0\,, \\
        \del{ \bar{m}^{(\Delta,\xi)}_n - \mathcal{M}^{(\Delta_3,\xi_3)}_{-n} - \mathcal{M}^{(\Delta_4,\xi_4)}_{-n}} \chi_{\Delta_{3,4},\xi_{3,4};\Delta,\xi}(t_{3},x_{3},t_4,x_4;\bar{v}) &=0\,, 
    \end{align}
\end{subequations}
for $n\geq-1$. Like for the Virasoro case, we have not found any closed-form solutions for the level-two wave functions. Nevertheless, in section \ref{sec:SemiClassBlocks} we make use of the semi-classical limit in order to approximate solutions for the above set of equations.

Finally, $\bms_3$-blocks are given by
\begin{equation}\label{4ptbmsWF}
\begin{split}
\langle0|\Op_{\Delta_4,\xi_4}&(t_4,x_4)\Op_{\Delta_3,\xi_3}(t_3,x_3)\mathcal{P}_{\Delta,\xi}\Op_{\Delta_1,\xi_1}(t_1,x_1)\Op_{\Delta_2,\xi_2}(t_2,x_2)|0\rangle\\
    &= \int_{\mathbb{C}^\infty} [\dd^2 v]_{\Delta,\xi} \, \chi_{\Delta_{3,4},\xi_{3,4};\Delta,\xi}(t_{3},x_{3},t_4,x_4;\bar{v})\, \psi_{\Delta_{1,2},\xi_{1,2};\Delta,\xi}(t_{1},x_{1},t_2,x_2;v)\,,
\end{split}
\end{equation}
where $\mathcal{P}_{\Delta,\xi}$ is the projector onto the $\bms_3$ module $\mathscr{B}^{c_{\text{\tiny L}},c_{\text{\tiny M}}}_{\Delta,\xi}$, which when restricted to the module acts as the unit operator, and we have inserted a completeness relation. 

Without loss of generality we may consider the point configuration 
\begin{equation}\label{pointconfig}
    \{(t_i,x_i)\} = \{(t,x),(0,0),(1,0),(\infty,0) \} \,,
\end{equation}
such that the $\bms_3$-block $\mathcal{B}_{\Delta_\mathrm{tot},\xi_\mathrm{tot};\Delta,\xi}(t,x)$ is given by 
\begin{equation}\label{bmsblockINT1}
    \begin{split}
        &\mathcal{B}_{\Delta_\mathrm{tot},\xi_\mathrm{tot};\Delta,\xi}(t,x) \\ 
        &\quad= \lim_{\substack{t_4\to\infty \\ x_4\to0}} t_4^{2\Delta_4} e^{2\xi_4 \frac{x_4}{t_4}}\int_{\mathbb{C}^\infty} [\dd^2 v]_{\Delta,\xi} \, \chi_{\Delta_{3,4},\xi_{3,4};\Delta,\xi}(1,0,t_4,x_4;\bar{v}) \psi_{\Delta_{1,2},\xi_{1,2};\Delta,\xi}(t,x,0,0;v)\,,
    \end{split}
\end{equation}
where $\Delta_\mathrm{tot}=\{\Delta_1,\Delta_2,\Delta_3,\Delta_4\}$ and $\xi_\mathrm{tot}=\{\xi_1,\xi_2,\xi_3,\xi_4\}$. Furthermore, in the point configuration \eqref{pointconfig} the relationship between the wave function and its dual simplifies to 
\begin{equation}\label{wfTransPC}
    \chi_{\Delta_{3,4},\xi_{3,4};\Delta,\xi}(1,0,t_4,x_4;\bar{v}) =  t_4^{-2\Delta_4} \e{-2\xi_4 \frac{x_4}{t_4}} \overline{\psi_{\Delta_{3,4},\xi_{3,4};\Delta,\xi}(1,0,t_4^{-1},-x_4 t_4^{-2};{v})}\,.
\end{equation}
Plugging in the above relation into \eqref{bmsblockINT1} and implementing the limits, we reach the compact formula
\begin{equation}\label{bmsblockINT}
   \begin{split}
        \mathcal{B}_{\Delta_\mathrm{tot},\xi_\mathrm{tot};\Delta,\xi}&(t,x) =\int_{\mathbb{C}^\infty} [\dd^2 v]_{\Delta,\xi} \, \overline{\psi_{\Delta_{3,4},\xi_{3,4};\Delta,\xi}(1,0,0,0;{v})} \, \psi_{\Delta_{1,2},\xi_{1,2};\Delta,\xi}(t,x,0,0;v)\,.
   \end{split}
\end{equation}

The set of equations \eqref{diffeqpsilvl2BMS} with $n=0$ fixes the wave function to be of the form
\begin{align}\label{wfsolnk0BMS}
    \psi_{\Delta_{1,2},\xi_{1,2};\Delta,\xi}(t,x,0,0;v)&=t^{\Delta-\Delta_1-\Delta_2}\e{\frac{x}{t}\left(\xi-\xi_1-\xi_2\right)}F(\eta,\nu)\,,
\end{align}
where $F(\eta,\nu)$ is an unknown function, and we have introduced the combinations of oscillator- and spacetime variables
\begin{align}
\label{etanu}
    \eta_n=t^n v^{(1)}_n\,, \qquad \nu_n=n t^{n-1}x v^{(1)}_n +t^n v^{(2)}_n\,,
\end{align}
where $n\in\mathbb{N}$. The dual wave function follows from the transformation \eqref{wfTransPC}. In appendix \ref{sec:uniF} we prove that $F(\eta,\nu)$ is uniquely determined by the remaining differential equations.

%%%%%%%%%%%%%%%%%%%%%%%%%%%%%%%%%%%%%%%%%%%%%%%%%%%%%%%%%%%
%%%%%%%%%%%%%%%%%%%%%%%%%%%%%%%%%%%%%%%%%%%%%%%%%%%%%%%%%%%
%%%%%%%%%%%%%%%%%%%%%%%%%%%%%%%%%%%%%%%%%%%%%%%%%%%%%%%%%%%

\section{Semi-classical \texorpdfstring{$\bms_3$}{bms3}-blocks}\label{sec:SemiClassBlocks}
In this section we will apply the machinery of our oscillator construction of the highest-weight representation of $\bms_3$ in order to calculate $\bms_3$-blocks in the semi-classical limit $c_{\text{\tiny M}} \to \infty$, with $\frac{\Delta}{c_{\text{\tiny M}}}$, $\frac{\Delta_i}{c_{\text{\tiny M}}}$;  $\frac{\xi}{c_{\text{\tiny M}}}$, $\frac{\xi_i}{c_{\text{\tiny M}}}$, and $\frac{c_{\text{\tiny L}}}{c_{\text{\tiny M}}}$ kept fixed. Here $\Delta,\xi \in \mathscr{B}^{c_{\text{\tiny L}},c_{\text{\tiny M}}}_{\Delta,\xi}$ are the scaling dimension and rapidity of the internal primary operator of each block, while $\Delta_i$, $\xi_i$ with $i=1,..,4$ denote the scaling dimensions and rapidities of the external operators. In order to implement the semi-classical limit we express the affected quantities in terms of an auxiliary parameter $\mu$ as follows, 
\begin{equation}
    c_{\text{\tiny M}} = \mu^2 \tilde{c}_{\text{\tiny M}}\,, \quad c_{\text{\tiny L}} = \mu^2 \tilde{c}_{\text{\tiny L}}\,, \quad \Delta = \mu^2 \tilde{\Delta}\,, \quad \Delta_i = \mu^2 \tilde{\Delta}_i\,, \quad \xi = \mu^2 \tilde{\xi}\,, \quad \xi_i = \mu^2 \tilde{\xi}_i\,;
\end{equation}
hence the semi-classical limit corresponds to $\mu \to \infty$ while keeping the tilde-quantities fixed.\footnote{In the context of a flat-space holographic duality the central charge $c_{\text{\tiny M}}$ is dimensionful since it is dual to the inverse of Newton's constant. In the current treatment of the limit $c_{\text{\tiny M}}\to\infty$ we may assign $\tilde{c}_{\text{\tiny M}}$ to be dimensionful while $\mu$ is dimensionless. Let us also note that dimensionful quantities always appear in dimensionless products in the $\bms_3$-block.}

%%%%%%%%%%%%%%%%%%%%%%%%%%%%
%%%%%%%%%%%%%%%%%%%%%%%%%%%%

\subsection{Semi-classical Differential Equations}

The semi-classical limit allows us to evaluate the $\bms_3$-block, using the integral expression \eqref{bmsblockINT}, by means of the saddle-point approximation. In order to express the integrand of \eqref{bmsblockINT} in a form which is suitable for the saddle-point approximation we must consider the contribution due to the exponential function in the measure \eqref{eq:measure}; we rescale the oscillator variables $v_n^{(i)}$ and $\bar{v}^{(i)}_n$ as follows
\begin{equation}\label{v-rescaling}
    v^{(i)}_n \mapsto \mu v^{(i)}_n\,, \qquad \quad \bar{v}^{(i)}_n \mapsto \mu \bar{v}^{(i)}_n \,,
\end{equation}
which in turn motivates the introduction of the variables $\sigma_n$ and $\kappa_n$ defined by
\begin{equation}
    \eta_n = \mu \, \sigma_n\,, \qquad  \nu_n = \mu \, \kappa_n\,.
\end{equation}

In accordance with our intention to employ the saddle-point approximation, we introduce an exponential ansatz for the wave function \eqref{wfsolnk0BMS}
\begin{equation}\label{eq:expansatz}
    F(\sigma,\kappa) = \exp\left[\mu^2 S(\sigma,\kappa)\right]\,.
\end{equation}
In appendix \ref{sec:uniqS} we comment on the proof of uniqueness for the function $S(\sigma,\kappa)$ in the exponential ansatz.

Using \eqref{eq:expansatz} together with the expressions for $l_k$ given in \eqref{lkBMS}, $m_k$ given in \eqref{mkBMS}, and $\mathcal{L}_k$ and $\mathcal{M}_k$ given in \eqref{curlyLM}, the differential equations \eqref{psiBMSlvl2-lequation} and \eqref{psiBMSlvl2-mequation} for $n\geq1$ read, respectively,
\begin{subequations}\label{Seqs}
    \begin{align}
        \begin{split}
    0&=\sum_{n=1}^{\infty} n\left(\sigma_n\del{\partial_{\sigma_{k+n}}S-\partial_{\sigma_n}S}+\kappa_n\del{\partial_{\kappa_{k+n}}S- \partial_{\kappa_n}S}\right)\\&
    \quad-\frac{1}{4}\sum_{n=1}^{k-1} \partial_{\sigma_n}S\, \partial_{\kappa_{k-n}}S  + \tilde{A}_k\partial_{\sigma_k} S+ \tilde{B}_k\partial_{\kappa_k}S-\left(\tilde{\Delta}+k\tilde{\Delta}_1-\tilde{\Delta}_2\right)\,,
\end{split}\\
        0&=\sum_{n=1}^{\infty} n \sigma_n \del{\partial_{\kappa_{k+n}}S- \partial_{\kappa_n}S}-\frac{1}{8}\sum_{n=1}^{k-1} 
    \partial_{\kappa_{k-n}} S \, \partial_{\kappa_n} S + \tilde{A}_k\partial_{\kappa_k}S  - \del{\tilde{\xi}+k\tilde{\xi}_1-\tilde{\xi}_2} ,
    \end{align}
\end{subequations}
were we have defined
\begin{subequations}\label{AkBk-tildeGENERAL}
\begin{align}
    A_k &= \mu\del{-\frac{i}{2} \sqrt{2\tilde{\xi}-\frac{\tilde{c}_{\text{\tiny M}}}{12}} - k\sqrt{\frac{\tilde{c}_{\text{\tiny M}}}{48}}} \equiv \mu \cdot \tilde{A}_k \,, \\ 
    B_k &= \mu\del{i\frac{\tilde{c}_{\text{\tiny L}}-24\tilde{\Delta}}{48\sqrt{2\xi-\frac{\tilde{c}_{\text{\tiny M}}}{12}}}-  k\frac{\tilde{c}_{\text{\tiny L}}}{48\sqrt{\frac{\tilde{c}_{\text{\tiny M}}}{12}}}}\equiv\mu \cdot \tilde{B}_k\,. 
\end{align}
\end{subequations}
We arrived at the equations \eqref{Seqs} after dividing by a leading factor of $\mu^2$ and getting rid of the overall exponential function; the resulting terms proportional to $\frac{1}{\mu}$ are sub-leading in the limit $\mu\to\infty$ and have hence been dropped, leaving two first-order differential equations for $S(\sigma,\kappa)$.

%%%%%%%%%%%%%%%%%%%%%%%%%%%%%%
%%%%%%%%%%%%%%%%%%%%%%%%%%%%%%

\subsection{Proof of Exponentiation}\label{sec:exponentiation}
Our ansatz \eqref{eq:expansatz} has consequences for the general form of the $\bms_3$-block. Using the exponential ansatz \eqref{eq:expansatz} in the $n=0$ solution for the wave function \eqref{wfsolnk0BMS}; transforming the result using the wave function relation \eqref{lvl2wftransBMS}; returning back to the oscillator variables $v^{(i)}_n$, $\bar{v}^{(i)}_n$ and remembering the rescaling \eqref{v-rescaling}, the expression for the $\bms_3$-block given in \eqref{bmsblockINT} reads
\begin{equation}\label{bmsblockGen}
    \begin{split}
        &\mathcal{B}_{\Delta_\mathrm{tot},\xi_\mathrm{tot};\Delta,\xi}(t,x) \\ 
        &\sim t^{\Delta-\Delta_1-\Delta_2}e^{\frac{x}{t}\left(\xi-\xi_1-\xi_2\right)} \int_{\mathbb{C}^\infty} \del{\prod_{n=1}^{\infty} 16n^2 \mu^4 \, \dd^{2}v^{(1)}_{n}  \dd^{2}v^{(2)}_{n}} \exp\left[\mu^2\mathcal{I}(t,x;v,\bar{v}) \right],
    \end{split}
\end{equation}
where we have defined 
\begin{equation}\label{Idef}
    \mathcal{I}(t,x;v,\bar{v}) = -4\sum_{n=1}^{\infty}n\del{ v^{(1)}_{n}\bar{v}^{(2)}_{n}+v^{(2)}_{n}\bar{v}^{(1)}_{n}} + S(t,x;v) + \overline{S({v})}\,.
\end{equation}
Note that the oscillator variables in the sum are those which have been rescaled, providing the overall factor of $\mu^2$ in the exponential function.

In the semi-classical limit, i.e. when $\mu\to\infty$, the exponent of the exponential function in \eqref{bmsblockGen} is large and hence the integral is dominated by the stationary points of $\mathcal{I}(t,x;v,\bar{v})$; we may then use the saddle-point approximation in order to evaluate the integral. The stationary points $(w, \bar{w})$ are found by extremising $\mathcal{I}(t,x;v,\bar{v})$ and they satisfy 
\begin{subequations}\label{eq:saddlew}
\begin{align}
    {w}^{(1)}_m &= \frac{1}{4m}\frac{\partial \bar{S}}{\partial \bar{v}^{(2)}_m}\Bigg|_{\bar{v}^{(i)}=\bar{w}^{(i)}}\,,
        & {w}^{(2)}_m &= \frac{1}{4m}\frac{\partial \bar{S}}{\partial \bar{v}^{(1)}_m}\Bigg|_{\bar{v}^{(i)}=\bar{w}^{(i)}} \,, \\
     \bar{w}^{(1)}_m &= \frac{1}{4m}\frac{\partial S}{\partial v^{(2)}_m}\Bigg|_{v^{(i)}=w^{(i)}} \,,
        & \bar{w}^{(2)}_m &= \frac{1}{4m}\frac{\partial S}{\partial v^{(1)}_m}\Bigg|_{v^{(i)}=w^{(i)}} \,.
\end{align}
\end{subequations}
To approximate the $\bms_3$-block we plug the stationary points into the integrand of \eqref{bmsblockGen}, which gives\footnote{Generically there exits more than one stationary point. However, to keep the formula simple, we assume that there is only a single stationary point $(w, \bar{w})$.}
\begin{equation}\label{bmsblockSADDLE}
\begin{split}
    \mathcal{B}_{\Delta_\mathrm{tot},\xi_\mathrm{tot};\Delta,\xi}(t,x)\approx t^{\Delta-\Delta_1-\Delta_2}\exp\left[\frac{x}{t}\left(\xi-\xi_1-\xi_2\right)+\mu^2\mathcal{I}(t,x;w,\bar{w})\right]\,,
\end{split}
\end{equation}
where the overall factor $16n^2\mu^4$ in \eqref{bmsblockGen} cancels with the determinant of the Hessian evaluated at the stationary point. The factor $t^{\Delta-\Delta_1-\Delta_2}$ may be expressed as an exponential of a logarithm. We have thus shown that the $\bms_3$-block takes a unique exponential form in the semi-classical limit.

In the following subsections we present two illustrative examples for how to compute $\bms_3$-blocks using the oscillator construction. We will focus on $\bms_3$-blocks of the vacuum module $\mathscr{B}^{c_{\text{\tiny L}},c_{\text{\tiny M}}}_{0,0}$, namely the perturbatively heavy vacuum block and the heavy-light vacuum block. Our computations will take advantage of the saddle-point approximation presented in this subsection. 

%%%%%%%%%%%%%%%%%%%%%%%%%%%%%%%%%%%%%
%%%%%%%%%%%%%%%%%%%%%%%%%%%%%%%%%%%%%

\subsection{Perturbatively Heavy Vacuum \texorpdfstring{$\bms_3$}{bms3}-block}\label{sec:pertHeavy}

The perturbatively heavy vacuum $\bms_3$-block is defined by $\Delta=\xi=0$ as well as the external parameters $\Delta_i$ and $\xi_i$ being infinitesimal and of the same order $\epsilon$.\footnote{We set $\Delta=\xi=0$ by choosing $\tilde{\Delta}=\tilde{\xi}=0$ before taking $\mu\to\infty$.} As a simplifying assumption we pairwise identify the external operators such that $\Delta_1 = \Delta_2$, $\xi_1=\xi_2$ and $\Delta_3=\Delta_4$, $\xi_3=\xi_4$.

\subsubsection{Analytic Behaviour}\label{sec:analyticBEHAVIOUR}
As a consequence of setting $\xi=\Delta=0$ the quantities defined in equations \eqref{Ak_Bkfull} will have negative expressions under a square root; we remedy this issue by analytically continuing these quantities, resulting in the coefficients
\begin{subequations}\label{AkBk-pm}
\begin{align}
    {A}_k&={-}\sqrt{\frac{{c}_{\text{\tiny M}}}{48}} (k-1)\equiv A^+_k \,, &{B}_k &={-}\frac{ {c}_{\text{\tiny L}}}{2\sqrt{48 {c}_{\text{\tiny M}}}}\del{k- 1} \equiv B^+_k\,, \label{AkBk-pm+}\\
    \hat{A}_k&={-}\sqrt{\frac{{c}_{\text{\tiny M}}}{48}} (k+1)\equiv A^-_k \,, &\hat{B}_k &={-}\frac{ {c}_{\text{\tiny L}}}{2\sqrt{48  {c}_{\text{\tiny M}}}}\del{k+1} \equiv B^-_k\label{AkBk-pm-}\,, 
\end{align}
\end{subequations}
where we have chosen the branch $\sqrt{-1}=+i$. The absence of factors of $i$ in the above expressions means that the $\bms_3$ generators $l_n$ and $m_n$, and by extension the differential equations \eqref{Seqs}, are unaffected by the complex conjugation defined with the overline operation; hence the relation between the level-two wave functions \eqref{lvl2wftransBMS} is invalidated. This situation was alluded to below equations \eqref{Ak_Bkfull}. Under the current circumstances we have two independent wave functions which we introduce in a way that is consistent with equation \eqref{bmsblockINT}, i.e.
\begin{subequations}\label{psiPM}
\begin{align}
    \psi_{\Delta_{1},\xi_{1};0,0}(t,x,0,0;v) &\equiv \psi^+_{\Delta_{1},\xi_{1};0,0}(t,x,0,0;v) \,, \\ 
    \overline{\psi_{\Delta_{3},\xi_{3};0,0}(1,0,0,0;{v})} &\equiv {\psi^-_{\Delta_{3},\xi_{3};0,0}(1,0,0,0;\bar{v})}\,,
\end{align}
\end{subequations}
where the superscripts of the wave functions correlate with the superscripts of the coefficients \eqref{AkBk-pm} in their equations.\footnote{Although we will not use it here, the analytic continuation motivates an analogous notation for the $\bms_3$ generators, i.e. $l_n\equiv l^+_n$, $m_n\equiv m^+_n$ and $\bar{l}_n\equiv l^-_n$, $\bar{m}_n\equiv m^-_n$.} Subsequently, since the $n=0$ equations are independent of the coefficients \eqref{AkBk-pm}, we use the partial solution \eqref{wfsolnk0BMS} to immediately deduce 
\begin{subequations}
\begin{align}
    \psi^+_{\Delta_{1},\xi_{1};0,0}(t,x,0,0;v)&=t^{-2\Delta_1}\e{-2\xi_1\frac{x}{t}}F^+(\eta,\nu)\,,\\
    \psi^-_{\Delta_{3},\xi_{3};0,0}(1,0,0,0;\bar{v})&=F^-(\bar{\eta},\bar{\nu})\,,
\end{align}
\end{subequations}
and the exponential ansätze analogous to \eqref{eq:expansatz} take the form 
\begin{equation}\label{F+F-}
    F^+(\sigma,\kappa) = \exp\left[\mu^2 S^+(\sigma,\kappa)\right]\,, \qquad F^-(\bar{\sigma},\bar{\kappa}) = \exp\left[\mu^2 S^-(\bar{\sigma},\bar{\kappa})\right]\,.
\end{equation}
The two sets of differential equations for $S^+$ and $S^-$ must be considered separately.

%%%%%%%%%%%%%%%%%%%%%%%%%%%%%%%%%%%%%%%%%%%%%%%%%%%%%%%%%%%%%

\subsubsection{Solving the Differential Equations}\label{sec:VacuumDif}

The analysis leading up to the set of differential equations \eqref{Seqs} is general enough that the expressions may serve as equations for the wave functions \eqref{psiPM} if we make the appropriate substitutions. For $\psi^+_{\Delta_{1},\xi_{1};0,0}(t,x,0,0;v)$ we get
\begin{subequations}\label{eq:S+}
\begin{align}
    \begin{split}
    0&=\sum_{n=1}^{\infty} n\left(\sigma_n\del{\partial_{\sigma_{k+n}}S^+-\partial_{\sigma_n}S^+}+\kappa_n\del{\partial_{\kappa_{k+n}}S^+- \partial_{\kappa_n}S^+}\right)\\
    &\quad -\frac{1}{4}\sum_{n=1}^{k-1} \partial_{\sigma_n}S^+\, \partial_{\kappa_{k-n}}S^+ + \tilde{A}^+_k\partial_{\sigma_k}S^+ + \tilde{B}^+_k\partial_{\kappa_k}S^+-\tilde{\Delta}_1(k-1)\,,
\end{split}\\
    0&=\sum_{n=1}^{\infty} n \sigma_n \del{\partial_{\kappa_{k+n}}S^+- \partial_{\kappa_n}S^+}-\frac{1}{8}\sum_{n=1}^{k-1} \partial_{\kappa_{k-n}} S^+ \, \partial_{\kappa_n} S^+ + \tilde{A}^+_k\partial_{\kappa_k}S^+  -\tilde{\xi}_1(k-1) \,; 
\end{align}
\end{subequations}
and for $\psi^-_{\Delta_{3},\xi_{3};\Delta,\xi}(1,0,0,0;\bar{v})$ we get
\begin{subequations}\label{S^-diffeqALL}
\begin{align}
\begin{split}
\label{S^-_diffeqn2}
    0&=\sum_{n=1}^{\infty} n\left(\bar{\sigma}_n\del{\partial_{\bar{\sigma}_{k+n}}S^- -\partial_{\bar{\sigma}_n}S^-}+\bar{\kappa}_n\del{\partial_{\bar{\kappa}_{k+n}}S^- - \partial_{\bar{\kappa}_n}S^-}\right)\\&
    \quad -\frac{1}{4}\sum_{n=1}^{k-1} \partial_{\bar{\sigma}_n}S^-\, \partial_{\bar{\kappa}_{k-n}}S^- + \tilde{A}^-_k\partial_{\bar{\sigma}_k}S^- + \tilde{B}^-_k\partial_{\bar{\kappa}_k}S^- -\tilde{\Delta}_3(k-1)\,,
\end{split}\\
\label{S^-_diffeqn1}
    0&=\sum_{n=1}^{\infty} n \bar{\sigma}_n \del{\partial_{\bar{\kappa}_{k+n}}S^- - \partial_{\bar{\kappa}_n}S^-}-\frac{1}{8}\sum_{n=1}^{k-1} \partial_{\bar{\kappa}_{k-n}} S^- \, \partial_{\bar{\kappa}_n} S^- + \tilde{A}^-_k\partial_{\bar{\kappa}_k}S^-  -\tilde{\xi}_3(k-1) \,, 
\end{align}
\end{subequations}
where the tilde references the dependence on fixed quantities, as in equations \eqref{AkBk-tildeGENERAL}. 

$S^+$ and $S^-$ may be treated as expansions in the infinitesimal parameters $\tilde{\Delta}_1$, $\tilde{\xi}_1$ and $\tilde{\Delta}_3$, $\tilde{\xi}_3$, respectively. We choose $S^+$ and $S^-$ to be of leading order $\epsilon.$\footnote{In principle the leading terms of $S^+(\sigma,\kappa)$ and $S^-(\bar\sigma,\bar\kappa)$ may be constant functions of order one. Such constant functions may be dropped since they only contribute to the $\bms_3$-block as inconsequential multiplicative factors.} The above set of differential equations has the leading order $\epsilon$; hence the terms quadratic in derivatives of $S^\pm$ are sub-leading of order $\epsilon^2$ and may be dropped. 

For the expansions of $S^\pm$ which we are assuming, the stationary point coordinates \eqref{eq:saddlew} have the leading order $\epsilon$. This means that products of oscillator variables with $S^\pm$ give rise to terms which are sub-leading when evaluated at the stationary point -- such terms will be dropped, too. This reasoning leaves us with the sets of linear differential equations
\begin{subequations}
\begin{align}
    0&=\tilde{A}^+_k\partial_{\sigma_k}S^+ + \tilde{B}^+_k\partial_{\kappa_k}S^+-\tilde{\Delta}_1(k-1) +{\Op\big(\epsilon^2\big)}\,, \label{S+l} \\
    0&=  \tilde{A}^+_k\partial_{\kappa_k}S^+  -\tilde{\xi}_1(k-1) +{\Op\big(\epsilon^2\big)}\, \label{S+m} ;
\end{align}
\end{subequations}
as well as
\begin{subequations}
\begin{align}
    0&= \tilde{A}^-_k\partial_{\bar{\sigma}_k}S^- + \tilde{B}^-_k\partial_{\bar{\kappa}_k}S^- -\tilde{\Delta}_3(k-1) +{\Op\big(\epsilon^2\big)}\,, \label{S-l}\\
    0&= \tilde{A}^-_k\partial_{\bar{\kappa}_k}S^-  -\tilde{\xi}_3(k-1) +{\Op\big(\epsilon^2\big)}\,,\label{S-m}
\end{align}
\end{subequations}
where the omitted terms are indicated by their order at the stationary point. Plugging the relevant expressions into the differential equations \eqref{S+m} and \eqref{S-m}, and integrating, we get
\begin{subequations}
\begin{align}
        S^+({\sigma},{\kappa}) &= -\sqrt{\frac{48}{\tilde{c}_{\text{\tiny M}}}}\tilde{\xi}_1 \sum_{n=1}^\infty{\kappa}_n + f({\sigma}) +{\Op\big(\epsilon^3\big)}\,,\\
        S^-(\bar{\sigma}, \bar{\kappa}) &= -\sqrt{\frac{48}{\tilde{c}_{\text{\tiny M}}} }\tilde{\xi}_3\sum_{n=1}^{\infty} \frac{n-1}{n+1}  \bar{\kappa}_n + g(\bar{\sigma}) +{\Op\big(\epsilon^3\big)}\,,
\end{align}
\end{subequations}
where $f(\sigma)$ and $g(\bar{\sigma})$ arise as constants of integration. The functions $f(\sigma)$ and $g(\bar{\sigma})$ are determined by integrating the equations \eqref{S+l} and \eqref{S-l} after the insertion of the above results; we get 
\begin{subequations}
\begin{align}
    f(\sigma) &= -\sqrt{\frac{48}{\tilde{c}_{\text{\tiny M}}}}\del{\tilde{\Delta}_1 - \tilde{\xi}_1\frac{\tilde{c}_{\text{\tiny L}}}{2\tilde{c}_{\text{\tiny M}}}}\sum_{n=1}^\infty\sigma_n +{\Op\big(\epsilon^3\big)}\, , \\
     g(\bar{\sigma}) &= -\sqrt{\frac{48}{\tilde{c}_{\text{\tiny M}}}}\del{\tilde{\Delta}_3 - \tilde{\xi}_3\frac{\tilde{c}_{\text{\tiny L}}}{2\tilde{c}_{\text{\tiny M}}}}\sum_{n=1}^\infty \frac{n-1}{n+1} \bar{\sigma}_n +{\Op\big(\epsilon^3\big)}\,.
\end{align}
\end{subequations}
Using  the variable transformations \eqref{etanu} and the rescalings \eqref{v-rescaling}, the leading order solutions for $S^\pm$ read
\begin{subequations}\label{Ssolnvac1}
\begin{align}
        S^+(t,x,0,0; {v}) &\approx -\sqrt{\frac{48}{\tilde{c}_{\text{\tiny M}}}} \left[\del{\tilde{\Delta}_1  - \tilde{\xi}_1\frac{\tilde{c}_{\text{\tiny L}}}{2\tilde{c}_{\text{\tiny M}}}}\sum_{n=1}^\infty t^n {v}^{(1)}_n +  \tilde{\xi}_1\sum_{n=1}^{\infty} \del{n t^{n-1} x {v}^{(1)}_n + t^n {v}^{(2)}_n }  \right] \,,\label{+vacuumSoln}\\
        S^-(1,0,0,0;\bar{v}) &\approx -\sqrt{\frac{48}{\tilde{c}_{\text{\tiny M}}}} \left[\del{\tilde{\Delta}_3 - \tilde{\xi}_3\frac{\tilde{c}_{\text{\tiny L}}}{2\tilde{c}_{\text{\tiny M}}}}\sum_{n=1}^\infty  \frac{n-1}{n+1} \bar{v}^{(1)}_n + \tilde{\xi}_3\sum_{n=1}^{\infty}  \frac{n-1}{n+1} \bar{v}^{(2)}_n\right]  \, .\label{S-vacuumSoln} 
\end{align}
\end{subequations}
Note that our solutions for $S^+$ and $S^-$ are true up to order $\epsilon^2$ when evaluated at the saddle point.

%%%%%%%%%%%%%%%%%%%%%%%%%%%%%%%%%%%%%%%%

\subsubsection{Implementing the Saddle-point Approximation}

Evaluating the $\bms_3$-block according to the formula \eqref{bmsblockSADDLE} requires us to determine $\mathcal{I}(t,x;w,\bar{w})$ at the stationary point $(w,\bar{w})$. Appropriately substituting the solutions \eqref{Ssolnvac1} into the formulae \eqref{eq:saddlew} we find that the coordinates of the stationary point are 
\begin{subequations}\label{saddlepointsVac1}
\begin{align}
    {w}^{(1)}_m &= \frac{1}{4m}\frac{\partial S^-}{\partial \bar{v}^{(2)}_m} = -\frac{1}{4m}\sqrt{\frac{48}{\tilde{c}_{\text{\tiny M}}}}  \frac{m-1}{m+1} \tilde{\xi}_3\,,\label{w1m}\\
    {w}^{(2)}_m &= \frac{1}{4m}\frac{\partial S^-}{\partial \bar{v}^{(1)}_m} = -\frac{1}{4m}\sqrt{\frac{48}{\tilde{c}_{\text{\tiny M}}}} \frac{m-1}{m+1} \del{ \tilde{\Delta}_3 - \tilde{\xi}_3 \frac{\tilde{c}_{\text{\tiny L}}}{2 \tilde{c}_{\text{\tiny M}}} }\,, \label{w2m}\\
    \bar{w}^{(1)}_m &= \frac{1}{4m}\frac{\partial S^+}{\partial v^{(2)}_m} = -\frac{1}{4m} \sqrt{\frac{48}{\tilde{c}_{\text{\tiny M}}}} \tilde{\xi}_1 t^m \,, \label{wbar1m} \\
    \bar{w}^{(2)}_m &= \frac{1}{4m}\frac{\partial S^+}{\partial v^{(1)}_m} = -\frac{1}{4m} \sqrt{\frac{48}{\tilde{c}_{\text{\tiny M}}}} t^m \left({ \tilde{\Delta}_1 - \tilde{\xi}_1 \frac{\tilde{c}_{\text{\tiny L}}}{2 \tilde{c}_{\text{\tiny M}}} } + \tilde{\xi}_1 m\frac{x}{t} \right) \label{wbar2m} \,.
\end{align}
\end{subequations}
It is practical to consider the terms of $\mathcal{I}(t,x;w,\bar{w})$, as defined in equation \eqref{Idef}, separately. Plugging the values \eqref{saddlepointsVac1} into the sum of oscillator variables yields  
\begin{align}\label{saddlesum1}
\begin{split}
    -4 &\sum_{n=1}^{\infty}n\del{ w^{(1)}_{n}\bar{w}^{(2)}_{n}+w^{(2)}_{n}\bar{w}^{(1)}_{n}} = -\frac{2}{\tilde{c}_{\text{\tiny M}}} \left[ \del{\tilde{\Delta}_3 \tilde{\xi}_1 + \tilde{\Delta}_1 \tilde{\xi}_3- \tilde{\xi}_3 \tilde{\xi}_1 \frac{\tilde{c}_{\text{\tiny L}}}{\tilde{c}_{\text{\tiny M}}}} \mathcal{F}(t) + \tilde{\xi}_3 \tilde{\xi}_1 x \partial_t \mathcal{F}(t) \right]\,,
\end{split}
\end{align}
with the definition
\begin{equation}\label{hypergeometricsum}
     \sum_{n=2}^\infty \frac{n-1}{n(n+1)}t^n = \frac{t^2}{6}\, \tensor[_2]{F}{_1}(2,2;4;t)\equiv\frac{1}{6}\mathcal{F}(t)\,,
\end{equation}
where $\tensor[_2]{F}{_1}(2,2;4;t)$ is a hypergeometric function. Moreover, $\mathcal{F}(t)$ satisfies the identity $\mathcal{F}(t)=6\left(\frac{t-2}{t}\ln(1-t)-2\right).$

The functions $S^+(t,x;v)$ and $S^-(\bar{v})$ coincide at the stationary point, i.e. evaluating them at the coordinates \eqref{saddlepointsVac1} gives the same result for both, 
\begin{equation}
    \begin{split}
        S^+(t,x;w) = S^-(\bar{w}) =\frac{2}{\tilde{c}_{\text{\tiny M}}} \left[ \del{\tilde{\Delta}_3 \tilde{\xi}_1 + \tilde{\Delta}_1 \tilde{\xi}_3- \tilde{\xi}_3 \tilde{\xi}_1 \frac{\tilde{c}_{\text{\tiny L}}}{\tilde{c}_{\text{\tiny M}}}} \mathcal{F}(t) + \tilde{\xi}_3 \tilde{\xi}_1 x \partial_t \mathcal{F}(t) \right]\,,
    \end{split}
\end{equation}
which follows from the same identifications as for the expression \eqref{saddlesum1}.

Using the above results in the definition
\begin{equation}\label{Iw}
\begin{split}
    \mathcal{I}(t,x;w,\bar{w}) = -4\sum_{n=1}^{\infty}n\del{ w^{(1)}_{n}\bar{w}^{(2)}_{n}+w^{(2)}_{n}\bar{w}^{(1)}_{n}} + S^+(t,x;w) + S^-(\bar{w})\,,
\end{split}
\end{equation}
and plugging the expression into the saddle-point approximation \eqref{bmsblockSADDLE} brings us to the expression for the perturbatively heavy vacuum $\bms_3$-block 
\begin{equation}\label{bmsblockVacSADDLE}
\begin{split}
    \mathcal{B}&_{\Delta_{1,3},\xi_{1,3};0,0}(t,x) \\
    &\approx t^{-2\Delta_1}\exp\left[ -2\xi_1\frac{x}{t}+\frac{2}{{c}_{\text{\tiny M}}} \left( \del{{\Delta}_3 {\xi}_1 + {\Delta}_1 {\xi}_3- {\xi}_3 {\xi}_1 \frac{{c}_{\text{\tiny L}}}{{c}_{\text{\tiny M}}}} \mathcal{F}(t) + {\xi}_3 {\xi}_1 x \partial_t \mathcal{F}(t) \right) \right]\,,
\end{split}
\end{equation}
where factors of $\mu$ have been absorbed into the non-tilde quantities. 

%%%%%%%%%%%%%%%%%%%%%%%%%%%%%%%%%%%%%%
%%%%%%%%%%%%%%%%%%%%%%%%%%%%%%%%%%%%%%

\subsection{Heavy-light Vacuum \texorpdfstring{$\bms_3$}{bms3}-block}\label{subsec:HeavyLight}
The heavy-light vacuum $\bms_3$-block has two heavy and two light external operators; the light operators have infinitesimal scaling dimensions and rapidities of order $\epsilon$, while those of the heavy operators are of order one.\footnote{Other authors define light operators to scale as order $(c_{\text{\tiny M}})^0$ in the semi-classical limit $c_{\text{\tiny M}} \rightarrow \infty$. Hence a perhaps more correct denomination would be to call the operators under consideration `perturbatively heavy'.} Continuing with the pairings used in the previous subsection, we assign $\Delta_1$, $\xi_1$ to the light operators and $\Delta_3$, $\xi_3$ to the heavy operators; hence, in the semi-classical limit $\tilde{\Delta}_1$, $\tilde{\xi}_1$ are of infinitesimal order $\epsilon$ while $\tilde{\Delta}_3$, $\tilde{\xi}_3$ are of order one. The vacuum block dictates that $\Delta=\xi=0$.

%%%%%%%%%%%%%%%%%%%%%%%%%%%%%%%%%%%%%%%%%%%%%%%%%%%5

\subsubsection{Analysing the Differential Equations}
The prescription for determining the exponents of the semi-classical wave-function ansätze \eqref{F+F-} is analogous to the procedure in subsection \ref{sec:VacuumDif}. In fact, the properties of the equations containing $\tilde{\Delta}_1$ and $\tilde{\xi}_1$ are unchanged and hence the solution for $S^+$ is identical to the expression \eqref{+vacuumSoln}; thus also the coordinates of the stationary point \eqref{wbar1m} and \eqref{wbar2m} are the same and of infinitesimal order. However, since $\tilde{\Delta}_3$ and $\tilde{\xi}_3$ are of finite order the discussion below equations \eqref{S^-diffeqALL} no longer applies to $S^-$; we must thus re-consider its differential equations and solution.

We motivate an appropriate ansatz for $S^-(\bar{\sigma},\bar{\kappa})$ by analysing the lowest non-trivial infinitesimal order of $\mathcal{I}(t,x;w,\bar{w})$, as given in \eqref{Iw}. As mentioned above, the stationary-point coordinates $\bar{w}^{(i)}_m$ are of infinitesimal order $\epsilon$; the coordinates ${w}^{(i)}_m$ are of order one, hence the contribution of the measure to $\mathcal{I}(t,x;w,\bar{w})$ is of order $\epsilon$. Similar argumentation leads to the conclusion that $S^+(t,x;w)$ is of order $\epsilon$, too. Hence, any terms of higher order than $\epsilon$ which arise from $S^-(\bar{w})$ are sub-leading. Since the variables $\bar{\kappa}_n$ and $\bar{\sigma}_n$ are of order $\epsilon$ when evaluated at the stationary point, we conclude that $S^-(\bar{\sigma},\bar{\kappa})$ should be at most linear in the variables and thus a suitable ansatz is\footnote{As before, we may ignore any constant terms in the ansatz for $S^-$ since those only contribute as multiplicative factors to the $\bms_3$-block.}
\begin{equation}\label{eq:ansatzSplus}
    S^{-}(\bar{\sigma}, \bar{\kappa})=\sum_{n=1}^{\infty} {C}_n \bar{\sigma}_n+\sum_{n=1}^{\infty}{D}_n \bar{\kappa}_n\,,
\end{equation}
where the coefficients $C_n$ and $D_n$ depend on $\tilde{\xi}_3$, $\tilde{\Delta}_3$. Plugging the above ansatz into the differential equations \eqref{eq:S+} yields 
\begin{subequations}
\begin{align}
    \begin{split}
    0&=\sum_{n=1}^{\infty} n\left(\bar{\sigma}_n\del{C_{k+n}-C_n}+\bar{\kappa}_n\del{D_{k+n} - D_n}\right)-\frac{1}{4}\sum_{n=1}^{k-1} C_n D_{k-n}\\
    &\quad + \tilde{A}^-_k C_k + \tilde{B}^-_k D_k-\tilde{\Delta}_3(k-1)\,,
\end{split}\\
    0&=\sum_{n=1}^{\infty} n \bar{\sigma}_n \del{D_{k+n}-D_{n}}-\frac{1}{8}\sum_{n=1}^{k-1} D_{k-n}D_n + \tilde{A}^-_k D_k -\tilde{\xi}_3(k-1) \,.
\end{align}
\end{subequations}
The leading terms of the above equations are of order one. Invoking the perspective of the saddle-point approximation, the variables $\bar{\sigma}_n$, $\bar{\kappa}_n$ give rise to sub-leading terms of order $\epsilon$; thus terms containing these variables may be dropped. The remaining set of equations takes the form of recurrence relations, i.e. 
\begin{subequations}\label{recrel1}
\begin{align}
    0&=-\frac{1}{4}\sum_{n=1}^{k-1} C_n D_{k-n} -(k+1)\sqrt{\frac{\tilde{c}_{\text{\tiny M}}}{48}} C_k -(k+1)\frac{\tilde{c}_{\text{\tiny L}}}{2\sqrt{48\tilde{c}_{\text{\tiny M}}}} D_k-\tilde{\Delta}_3(k-1)+{\Op\big(\epsilon\big)}\,,\\
    0&=-\frac{1}{8}\sum_{n=1}^{k-1} D_{k-n}D_n -(k+1)\sqrt{\frac{\tilde{c}_{\text{\tiny M}}}{48}} D_k -\tilde{\xi}_3(k-1) +{\Op\big(\epsilon\big)} \,,
\end{align}
\end{subequations}
where the omitted terms are indicated by their order at the stationary point $(w,\bar{w})$, and we used the definitions \eqref{AkBk-pm-} for $\tilde{A}^-_k$ and $\tilde{B}^-_k$. Note that for $k=1$ the above recurrence relations fix the initial values $C_1 = D_1=0$.

The recurrence relations \eqref{recrel1} may be turned into differential equations by using the method of generating functions. To this end, we make the ansätze
\begin{equation}\label{CDdef}
    C(\tau)=\sum\limits_{n=1}^{\infty}C_n\tau^n\quad\text{and}\quad D(\tau)=\sum\limits_{n=1}^{\infty}D_n\tau^n\,.
\end{equation}
As a consequence of the above ansätze we have $C(0)=D(0)=0$ and the constraints $C_1=D_1=0$ become the boundary conditions $C'(0)=0=D'(0)=0$.    

Using standard techniques, the recurrence relations \eqref{recrel1} are transformed into two coupled differential equations for $C(\tau)$ and $D(\tau)$, 
\begin{subequations}
\begin{align}
     \partial_{\tau}\left(\tau\cdot C(\tau)\right)&=-\sqrt{\frac{3}{\tilde{c}_{\text{\tiny M}}}} C(\tau)D(\tau)-\sqrt{\frac{48}{\tilde{c}_{\text{\tiny M}}}}\frac{\tilde{\Delta}_3 \tau^2}{(1-\tau)^2}-\frac{\tilde{c}_{\text{\tiny L}}}{2\tilde{c}_{\text{\tiny M}}}\partial_{\tau}\left(\tau\cdot D(\tau)\right)+{\Op\big(\epsilon\big)}\,, \\
    \partial_{\tau}\left(\tau\cdot D(\tau)\right)&=-\frac{1}{8}\sqrt{\frac{48}{\tilde{c}_{\text{\tiny M}}}}\left(D(\tau)\right)^2-\sqrt{\frac{48}{\tilde{c}_{\text{\tiny M}}}}\frac{\tilde{\xi}_3 \tau^2}{(1-\tau)^2} +{\Op\big(\epsilon\big)} \, .
\end{align}
\end{subequations}
We present the details of the solution procedure of the above differential equations in appendix \ref{apdx:differentialequations}; the solutions, to leading order in infinitesimal quantities, read
\begin{subequations}\label{CDsoln}
\begin{align}
\begin{split}\label{C_soln}
     C(\tau)&\approx-\frac{\tau(1-\tau)^{\beta_3-1}\left(24\tilde{\Delta}_3+\tilde{c}_{\text{\tiny L}}(\beta_3^2-1)\right)}{\sqrt{3\tilde{c}_{\text{\tiny M}}}\left((1-\tau)^{\beta_3}-1\right)^2}\ln(1-\tau)\\
    &\quad -\frac{12\tilde{\Delta}_3 \tau}{\sqrt{3\tilde{c}_{\text{\tiny M}}}\beta_3}\left(\frac{(1-\tau)^{\beta_3}+1}{(\tau-1)\left((1-\tau)^{\beta_3}-1\right)}\right)\\
    & \quad-\frac{\tilde{c}_{\text{\tiny L}}}{2\sqrt{3\tilde{c}_{\text{\tiny M}}}(\tau-1)}\left(\tau-2-\frac{\tau}{\beta_3}\frac{(1-\tau)^{\beta_3}+1}{\left((1-\tau)^{\beta_3}-1\right)}\right)\,,
\end{split}\\
     D(\tau)&\approx-\sqrt{\frac{\tilde{c}_{\text{\tiny M}}}{3}}\frac{1}{1-\tau}\left(2-\tau+\beta_3\tau\left(1-\frac{2}{1-(1-\tau)^{\beta_3}}\right)\right)\,,
\end{align}
\end{subequations}
with  
\begin{equation}
    \beta_3=\sqrt{1-\tilde{\xi}_3\frac{24}{\tilde{c}_{\text{\tiny M}}}} \, .
\end{equation}
Although the ansatz \eqref{eq:ansatzSplus} is expressed in terms of the coefficients $C_m$ and $D_m$, which can be extracted from the functions $C(\tau)$ and $D(\tau)$, we may also continue using the solutions as they are given in \eqref{CDsoln}.

%%%%%%%%%%%%%%%%%%%%%%%%%%%%%%%%%%%%%%

\subsubsection{Implementing the Saddle-point Approximation}
Returning to the rescaled oscillator variables \eqref{v-rescaling} via \eqref{etanu}, and remembering the solution for $S^+$ given in equation \eqref{+vacuumSoln}, we have
\begin{subequations}\label{SsolnHHLL}
\begin{align}
    S^+(t,x,0,0; {v}) &\approx -\sqrt{\frac{48}{\tilde{c}_{\text{\tiny M}}}} \left[\del{\tilde{\Delta}_1  - \tilde{\xi}_1\frac{\tilde{c}_{\text{\tiny L}}}{2\tilde{c}_{\text{\tiny M}}}}\sum_{n=1}^\infty t^n {v}^{(1)}_n +  \tilde{\xi}_1\sum_{n=1}^{\infty} \del{n t^{n-1} x {v}^{(1)}_n + t^n {v}^{(2)}_n }  \right] \,,\\
    S^-(1,0,0,0;\bar{v}) &=\sum\limits_{n=2}^{\infty}\left(C_n\bar{v}^{(1)}_n+D_n \bar{v}^{(2)}_n \right)\, .
\end{align}
\end{subequations}
Note that the sum in $S^-$ starts at $n=2$. 

Being cognisant of the saddle-point approximation of the $\bms_3$-block, given in equation \eqref{bmsblockSADDLE}, the next step is to evaluate $\mathcal{I}(t,x;w,\bar{w})$. Making use of the expressions \eqref{SsolnHHLL}, the coordinates of the stationary point read 
\begin{subequations}\label{saddlepointsHHLL}
\begin{align}   
    {w}^{(1)}_m &= \frac{1}{4m} D_m\,, &{w}^{(2)}_m &=  \frac{1}{4m} C_m \,, \\
    \bar{w}^{(1)}_m &= -\frac{1}{4m} \sqrt{\frac{48}{\tilde{c}_{\text{\tiny M}}}}  \tilde{\xi}_1 t^m \,, 
    &\bar{w}^{(2)}_m &= -\frac{1}{4m} \sqrt{\frac{48}{\tilde{c}_{\text{\tiny M}}}} t^m \left({ \tilde{\Delta}_1 - \tilde{\xi}_1 \frac{\tilde{c}_\text{\tiny L}}{2 \tilde{c}_\text{\tiny M}} } + m\,\tilde{\xi}_1 \frac{x}{t} \right) \,.
\end{align}
\end{subequations}
Plugging the above values into equation \eqref{Iw} yields
\begin{equation}
\label{I_HHLL}
    \mathcal{I} (t,x;w,\bar{w})=-\sqrt{\frac{48}{\tilde{c}_{\text{\tiny M}}}}\sum\limits_{m=2}^{\infty}\frac{1}{4m}t^m\left(\tilde{\xi}_1 C_m+\tilde{\Delta}_1 D_m -\tilde{\xi}_1 \frac{\tilde{c}_{\text{\tiny L}}}{2\tilde{c}_{\text{\tiny M}}} D_m  +  m \,\tilde{\xi}_1 \frac{x}{t} D_m \right)\,.
\end{equation}
Recalling the definitions \eqref{CDdef} we may express the coefficients $C_m$ and $D_m$ as the integrals
\begin{equation}
    \sum_{m=1}^{\infty}\frac{1}{m}C_m t^m =\int_{0}^{t}\dd\tau\, \frac{C(\tau)}{\tau}  \quad\mathrm{and}\quad \sum_{m=1}^{\infty}\frac{1}{m} D_m t^m =\int_{0}^{t}\dd\tau \, \frac{D(\tau)}{\tau}\,;
\end{equation}
hence
\begin{equation}\label{IsaddleHHLL}
    \mathcal{I}(t,x;w,\bar{w})=-\frac{1}{4}\sqrt{\frac{48}{\tilde{c}_\text{\tiny M}}}\left(\tilde{\xi}_1\int_0^t \dd\tau \frac{C(\tau)}{\tau}+\tilde{\Delta}_1\int_0^t \dd\tau \frac{D(\tau)}{\tau}-\tilde{\xi}_1\frac{\tilde{c}_\text{\tiny L}}{2\tilde{c}_\text{\tiny M}}\int_0^t \dd\tau \frac{D(\tau)}{\tau}+\tilde{\xi}_1\frac{x}{t}D(t)\right)\,,
\end{equation}
where the last term follows directly from \eqref{CDdef}. After evaluating the integrals, for which the expressions are presented in equations \eqref{C-int} and \eqref{D-int}, and using \eqref{bmsblockSADDLE}, we arrive at
\begin{equation}\label{HHLLsoln}
\begin{split}
    \mathcal{B}^{\mathrm{HHLL}}_{\Delta_{1,3},\xi_{1,3};0,0}(t,x)\approx \left(\frac{(1-t)^{\beta_3-1}}{(1-(1-t)^{\beta_3})^2}\right)^{\Delta_1} \exp[-\xi_1x\frac{(1-t)^{\beta_3}(1+\beta_3)+\beta_3-1}{(1-t)(1-(1-t)^{\beta_3})}]& \\
     \times \exp[\frac{12 \xi_1 }{c_{\text{\tiny M}}\beta_3}\left(\xi_3\frac{c_{\text{\tiny L}}}{c_{\text{\tiny M}}}-\Delta_3\right)\left(\frac{1+(1-t)^{\beta_3}}{1-(1-t)^{\beta_3}}\right)\ln(1-t)]&\,,
\end{split}
\end{equation}
where factors of $\mu$ are absorbed by the non-tilde quantities. 

Choosing $\Delta_3$, $\xi_3$ to be infinitesimal; expanding $\beta_3$ to first order in $\xi_3$; and keeping terms up to second order in infinitesimal quantities, transforms the above result into the perturbatively heavy vacuum $\bms_3$-block (\ref{bmsblockVacSADDLE}), modulo constant prefactors. Furthermore, the result in \eqref{HHLLsoln} generalises the heavy-light vacuum $\mathfrak{bms}_3$-block presented in \cite{Hijano:2018nhq}.

%%%%%%%%%%%%%%%%%%%%%%%%%%%%%%%%%%%%%%%%%%%%%%%%%%%%%%%%%%%
%%%%%%%%%%%%%%%%%%%%%%%%%%%%%%%%%%%%%%%%%%%%%%%%%%%%%%%%%%%
%%%%%%%%%%%%%%%%%%%%%%%%%%%%%%%%%%%%%%%%%%%%%%%%%%%%%%%%%%%

\section{Summary and Outlook}\label{sec:conclusions}
In this paper we presented an oscillator construction of highest-weight $\mathfrak{bms}_3$ modules. We used this construction to obtain general expressions for two- and three-point correlation functions and the $\mathfrak{bms}_3$-block in terms of wave functions, where the wave functions must satisfy an infinite set of partial differential equations. We found an expression for the level-one wave function and its dual -- we could not, however, obtain closed-form solutions for the level-two wave functions; nevertheless, we proved that they are uniquely determined in generic cases.

We showed the strengths of the oscillator construction by considering the semi-classical limit defined by $c_{\text{\tiny M}} \rightarrow \infty$ while keeping all fractions of the form $\frac{\Delta}{c_{\text{\tiny M}}}$, $\frac{\Delta_i}{c_{\text{\tiny M}}}$; $\frac{\xi}{c_{\text{\tiny M}}}$, $\frac{\xi_i}{c_{\text{\tiny M}}}$ and $\frac{c_{\text{\tiny L}}}{c_{\text{\tiny M}}}$ fixed. Our oscillator construction allowed us to prove the exponentiation of $\mathfrak{bms}_3$-blocks in this limit. Finally, we explicitly determined the vacuum $\mathfrak{bms}_3$-block in the perturbatively heavy and the heavy-light regimes; the heavy-light vacuum $\bms_3$-block is in agreement with previous results stemming from holographic computations involving probe particles propagating in flat-space cosmologies within Einstein gravity \cite{Hijano:2018nhq}. It would be interesting to check our predictions against holographic calculations in the semi-classical limit of chiral gravity theories with $c_\text{\tiny L} \neq 0$.

The work presented in this paper allows for a number of possible directions which may be explored in the future. 

In our proof of uniqueness of the level-two wave functions we could identify a discrete set of values for $c_\text{\tiny L}$, $c_\text{\tiny M}$, $\Delta$ and $\xi$ which must be excluded. Analysing the analogous proof for the Virasoro algebra \cite{Besken:2019jyw}, we notice that the set of excluded parameters in this case include the known values of Virasoro minimal models. Following this reasoning, we speculate that this method may be of use for identifying new $\bms_3$ minimal models.

It would be interesting to study $\bms_3$-blocks beyond the vacuum module, as well as blocks in which all external operators are heavy. In addition, $\frac{1}{c_\text{\tiny M}}$-corrections to the semi-classical limit are of interest since they give rise to quantum contributions to the bulk gravitational theory. Another avenue worth exploring is the computation of torus $\bms_3$-blocks in the semi-classical limit; we suspect that the oscillator construction would be beneficial for this task. The study of one-point functions on the torus was recently initiated in \cite{Bagchi:2019unf,Bagchi:2020rwb}. We expect that modularity gives similar stringent constraints on the thermodynamics of the dual gravitational theory as in the case of asymptotically-AdS spacetimes.

Furthermore, it would be valuable to have access to an oscillator construction for induced representations of $\bms_3$, since these representations are unitary \cite{Barnich:2014kra,Campoleoni:2016vsh}. As a first step in this direction one could approach an oscillator construction of induced $\isl_2$ modules. Moreover, the oscillator construction exhibits similarities with the approach of coadjoint orbits; see \cite{Taylor:1993zp} for insights in the case of the Virasoro group. Since coadjoint orbits are related to unitary representations \cite{Witten:1987ty} developing this relationship may provide an effective angle of approach.

In view of the recent results of \cite{Kapec:2014opa,Kapec:2016jld,Pasterski:2016qvg,Duval:2014uva}, it would also be interesting to obtain an oscillator construction of $\bms_4$ algebras and investigate its relation to conformal field theories on the celestial sphere.

In conclusion, all of the above tasks are of importance in view of a putative holographic duality involving asymptotically flat spacetimes and BMS-invariant field theories. We hope to return to some of these aspects in future work.

%%%%%%%%%%%%%%%%%%%%%%%%%%%%%%%%%%%%%%%%%%%%%%%%%%%%%%%%%%%
%%%%%%%%%%%%%%%%%%%%%%%%%%%%%%%%%%%%%%%%%%%%%%%%%%%%%%%%%%%
%%%%%%%%%%%%%%%%%%%%%%%%%%%%%%%%%%%%%%%%%%%%%%%%%%%%%%%%%%%

\section*{Acknowledgements}
The work of SG, CM, MP and KW is funded by the \emph{Deutsche Forschungsgemeinschaft (DFG)} under Grant No.\,406116891 within the Research Training Group RTG\,2522/1. MA is funded by the \emph{Deutsche Forschungsgemeinschaft} (DFG, German Research  Foundation) under Grant No.\,406235073 within the Heisenberg program. MP is funded by a \emph{Landesgraduiertenstipendium} of the federal state of Thuringia.
\appendix
\section{Realising the Oscillatory Construction of \texorpdfstring{$\bms_3$}{bms3}}\label{app:building_oscill_rep}
In section \ref{sec:OscConstr_bms} of the main text we state that the oscillator construction of the highest-weight representation of $\bms_3$ may be reached by taking a non-relativistic limit of a two-dimensional linear-dilaton like theory; in this section we detail the procedure.\footnote{The analogous procedure for the Virasoro case is reviewed in appendix A.2.1 of \cite{Besken:2019bsu}.} 

Generalising the linear dilaton theory, the components of the stress-tensor $T\equiv T_{zz}$ and $\bar{T}\equiv T_{\bar{z}\bar{z}}$, where the bar denotes quantities belonging to the anti-holomorphic sector of the two-dimensional conformal algebra, may be expressed as
\begin{equation}
    T = \sum_{m=-\infty}^\infty {L}^{\vir}_m z^{-m+2} \, , \qquad \textrm{with} \quad L^{\vir}_m = \frac{1}{2} \sum_{n=-\infty}^\infty : \alpha_{m-n} \alpha_n: + i (m+1) V \alpha_m
\end{equation}
as well as
\begin{equation}
    \bar{T} = \sum_{m=-\infty}^\infty \bar{{L}}^{\vir}_m \bar{z}^{-m+2} \, , \qquad \textrm{with} \quad \bar{{L}}^{\vir}_m = \frac{1}{2} \sum_{n=-\infty}^\infty : \bar{\alpha}_{m-n} \bar{\alpha}_n: + i (m+1) \bar{V} \bar{\alpha}_m\,,
\end{equation}
where $L^\vir_m$, $\bar{L}^\vir_m$ are generators of the Virasoro algebra, and $V$, $\bar{V}$ are complex constants which are not related by complex conjugation.\footnote{Note that there is no action principle associated to this kind of theory, which is why we use the wording `dilaton-like'.} We impose $\big(L^\vir_m\big)^\dagger = L^\vir_{-m}$ and $\big(\bar{L}^\vir_m\big)^\dagger = \bar{L}^\vir_{-m}$. The oscillators $\alpha_m$, $\bar{\alpha}_m$ satisfy the commutation relations
\begin{equation}\label{alphacommutation}
  [\alpha_m, \alpha_n] = [\bar{\alpha}_m, \bar{\alpha}_n]= m \, \delta_{m+n,0} \, , \qquad [\alpha_m,\bar{\alpha}_n]=0\,.
\end{equation}
We will take the non-relativistic limit at the level of the generators $L^\vir_m$ and $\bar{L}^\vir_m$.

\subsection{Non-relativistic Limit}
In order to implement the non-relativistic limit we introduce a new pair of oscillators
\begin{align}\label{betagamma}
        \beta_m = \frac{1}{\sqrt{\epsilon}} \del{\alpha_m -i\bar{\alpha}_m}\,,  \quad \gamma_m = {\sqrt{\epsilon}} \del{\alpha_m + i\bar{\alpha}_m}\,.
\end{align}
It follows from the commutation relations \eqref{alphacommutation} that the oscillators defined above satisfy
\begin{equation}\label{betagammacommutators}
 [\beta_m, \gamma_n] = 2m \delta_{m+n,0}\,, \qquad [\beta_m, \beta_n] = [\gamma_m,\gamma_n] = 0\,.    
\end{equation}
Furthermore, we introduce the new constants
\begin{equation}\label{WL_WM}
     W_{\text{\tiny L}} = \frac{1}{2\sqrt{\epsilon}}\del{V-i\bar{V}}\,, \quad W_{\text{\tiny M}}=\frac{\sqrt{\epsilon}}{2}\del{V+i\bar{V}}\,.
\end{equation}

Imposing the adjoint properties $\beta_m^\dagger = \beta_{-m}$ and $\gamma_m^\dagger = \gamma_{-m}$ means that the Virasoro oscillators transform like $\alpha_m^\dagger {=}\alpha_{-m}$ and $\bar{\alpha}_m^\dagger {=}-\bar{\alpha}_{-m}$ for $m\neq0$. Preserving the adjoint property of the Virasoro generators in turn requires $V^* = V$ and $\bar{V}^* = -\bar{V}$, together with $\alpha_0^\dagger = \alpha_0 + 2iV$ and $\bar{\alpha}_0^\dagger = -\bar{\alpha}_0 - 2i\bar{V}$.

The $\bms_3$ generators are reached by taking a non-relativistic limit of the relativistic Virasoro generators. In practice this is achieved by taking the linear combinations \cite{Bagchi:2009pe,Hosseiny:2009jj}
\begin{subequations}
\begin{align}
    L_m &\equiv \lim_{\epsilon\to0}\del{L^{\vir}_m + \bar{L}^{\vir}_m}\,, \\
    M_m &\equiv \lim_{\epsilon\to0}\epsilon\del{L^{\vir}_m - \bar{L}^{\vir}_m}\,.
\end{align}
\end{subequations}
Expressing $L^\vir_m$ and $\bar{L}^\vir_m$ in terms of the oscillators \eqref{betagamma}, applying the linear combinations above, and taking $\epsilon\to0$, gives
\begin{subequations}\label{bmsGenlimit}
\begin{align}
    L_m &= \frac{1}{4}\sum_{n=-\infty}^\infty  :\beta_{m-n} \gamma_{n} + \gamma_{m-n}\beta_n: + i (m+1) \del{W_{\text{\tiny L}} \gamma_m + W_{\text{\tiny M}} \beta_m}\,, \\
    M_m &= \frac{1}{4}\sum_{n=-\infty}^\infty  :\gamma_{m-n} \gamma_{n}: + i (m+1) W_{\text{\tiny M}} \gamma_m\,.
\end{align}
\end{subequations}
The generators \eqref{bmsGenlimit} satisfy the $\bms_3$ algebra \eqref{eq:bms3} with central charges $c_{\text{\tiny L}} = 2 + 48 W_{\text{\tiny L}} W_{\text{\tiny M}}$ and $c_{\text{\tiny M}} = 24 W_{\text{\tiny M}}^2$. Enforcing that $L_m^\dagger = L_{-m}$ and $M_m^\dagger = M_{-m}$ requires, in addition to the adjoint properties of $\beta_m$ and $\gamma_m$ for $m\neq0$, that $\beta_0^\dagger = \beta_0 + 4i W_{\text{\tiny L}}$ and $\gamma_0^\dagger = \gamma_0 + 4i W_{\text{\tiny M}}$.

It is advantageous to work with more familiar commutation relations than those in \eqref{betagammacommutators}. We introduce a third set of oscillators as 
\begin{equation}
    a_m = \frac{1}{2}  \left( \beta_m + \gamma_m \right) \, , \qquad \hat{a}_m = \frac{1}{2}  \left( \beta_m - \gamma_m \right)\, ,
\end{equation}
which, using \eqref{betagammacommutators}, satisfy 
\begin{equation}
    [a_m,a_n] =  m \delta_{m+n,0}\,, \qquad [\hat{a}_m,\hat{a}_n] =  -m \delta_{m+n,0}, \qquad [a_m,\hat{a}_n] = 0\,.
\end{equation}
In terms of the above definitions the $\bms_3$ generators \eqref{bmsGenlimit} read 
\begin{subequations}\label{LMbmsa}
\begin{align}
    L_m &=  \frac{1}{2} \sum_{n=-\infty}^\infty : a_{m-n} a_n - \hat{a}_{m-n} \hat{a}_n : + i(m+1) \left[\del{W_{\text{\tiny M}}+W_{\text{\tiny L}}} a_m + \del{W_{\text{\tiny M}}-W_{\text{\tiny L}}} \hat{a}_m   \right] \, ,\label{Lbmsa}  \\
    M_m &=  \frac{1}{4} \sum_{n=-\infty}^\infty : {a}_{m-n}{a}_n -({a}_{m-n}\hat{a}_n + \hat{a}_{m-n}{a}_n ) +\hat{a}_{m-n} \hat{a}_n : + i(m+1) W_{\text{\tiny M}} (a_m-\hat{a}_m) \, . \label{Mbmsa} 
\end{align}
\end{subequations}
Preserving the adjoint properties $L^\dagger_m = L_{-m}$ and $M^\dagger_m = M_{-m}$ requires $a^\dagger_m = a_{-m}$, ${a}^\dagger_m = \hat{a}_{-m}$ for $m\neq0$, as well as 
\begin{equation}\label{a0transf}
    a_0^\dagger = a_0 + 2i(W_{\text{\tiny M}}+W_{\text{\tiny L}}) \,,\qquad \hat{a}_0^\dagger = \hat{a}_0 - 2i (W_{\text{\tiny M}} - W_{\text{\tiny L}})\,.
\end{equation}
The eigenvalues of the oscillators $a_m$ and $\hat{a}_m$ are in general complex: the real parts of the eigenvalues are arbitrary and parametrised by $\lambda_1$ and $\lambda_2$, respectively; the imaginary parts are fixed by the above expressions. From the transformations \eqref{a0transf} we read off the eigenvalues 
\begin{equation}\label{a0lambdamu}
    a_0 \equiv \sqrt{2}\lambda_1 + i\sqrt{2}\mu_1\,, \qquad  \hat{a}_0 \equiv \sqrt{2}\lambda_2 + i\sqrt{2}\mu_2 \,,
\end{equation}
where the factors of $\sqrt{2}$ are due to a choice of normalisation, and we defined $\mu_1 = -(W_{\text{\tiny M}} + W_{\text{\tiny L}})/\sqrt{2}$ and $\mu_2 = (W_{\text{\tiny M}} -W_{\text{\tiny L}})/\sqrt{2}$.

%%%%%%%%%%%%%%%%%%%%%%%%%%%
%%%%%%%%%%%%%%%%%%%%%%%%%%%%

\subsection{Differential Operators and Measure}\label{app:subsec:operators_measure}
We set up the oscillator construction of the $\bms_3$ generators \eqref{bmsGenlimit} by assigning differential operators to the modes $a_m$ and $\hat{a}_m$; we do so in the following way:
\begin{subequations}
\begin{align}
    a_{m} &= \frac{i}{2\sqrt{2}}\del{ \partial_{v^{(1)}_m} +  \partial_{v^{(2)}_m} }\,, &a_{-m} &= -i m\sqrt{2}\del{ {v^{(1)}_m} +  {v^{(2)}_m} }\,, \label{adiff} \\
    \hat{a}_{m} &= \frac{i}{2\sqrt{2}}\del{ \partial_{v^{(1)}_m} -  \partial_{v^{(2)}_m} }\,, &\hat{a}_{-m} &= im\sqrt{2}\del{ {v^{(1)}_m} - {v^{(2)}_m} }\,. \label{ahatdiff}
\end{align}
\end{subequations}
The adjoint transformation properties $a^\dagger_m = a_{-m}$ and $\hat{a}^\dagger_m =\hat{a}_{-m}$ are implied by the Hermitian product
\begin{equation}\label{aahatinnerprod}
    \del{ f , g } = \int_{\mathbb{C}^\infty} [\mathrm{d}^2 v] \, \overline{f(v)} g(v)\,,
\end{equation}
with the measure 
\begin{equation}\label{measurea-ahat}
    [\dd^2 v]=\prod_{n=1}^{\infty}{16n^2}\exp\left[-4n\del{v^{(1)}_{n}\bar{v}^{(2)}_{n}+v^{(2)}_{n}\bar{v}^{(1)}_{n}}\right]\dd^{2}v^{(1)}_{n}\dd^{2}v^{(2)}_{n}\,,
\end{equation}
where the normalisation is such that $(\mathbf{1},\mathbf{1})=1$. 

The exponential factor of the measure is made plausible by introducing the linear combinations $q_m = v_m^{(1)} + v_m^{(2)}$ and $j_m = v_m^{(1)} - v_m^{(2)}$. Then $a_{-m}$ only contains $q_m$ and $a_{-m}$ only contains $\partial_{q_m}$; while $\hat{a}_{-m}$ and $\hat{a}_{m}$ contain $j_m$ and its derivative, respectively. The measure of the oscillator construction of the Virasoro algebra, given by \eqref{virasoromeasure}, corresponds to an assignment akin to \eqref{adiff} \cite{Besken:2019bsu}; hence, replacing $u_n$ with $q_n$, the transformation $a^\dagger_m = a_{-m}$ requires a measure with the exponential factor
\begin{equation}\label{measurea}
    \exp[-2n \del{v^{(1)}_n + v^{(2)}_n}\del{\bar{v}^{(1)}_n + \bar{v}^{(2)}_n}] \, .
\end{equation}
Taking the overall sign difference between $a_{-m}$ and $\hat{a}_{-m}$ into consideration, similar reasoning motivates the exponential factor that allows for $\hat{a}^\dagger_m =\hat{a}_{-m}$ to be of the form
\begin{equation}\label{measureahat}
    \exp[2n \del{v^{(1)}_n - v^{(2)}_n}\del{\bar{v}^{(1)}_n - \bar{v}^{(2)}_n}] \, .
\end{equation}
Taking the product of the factors \eqref{measurea} and \eqref{measureahat} we arrive at the exponential of the measure \eqref{measurea-ahat}.

Evaluating the integral in equation \eqref{aahatinnerprod} for a general function $f(v,\bar{v})$ is made possible by the analytic continuation associated with the mapping $n\mapsto in$. Such an analytic continuation allows us to exploit the behaviour of the complex delta distribution and its derivatives, defined by
\begin{align}\label{eq:complex_delta}
    \int_{\mathbb{C}}\!\!\d v \d\bar{v}\  f(v,\bar{v})\ \partial_{v}^{\,a}\partial_{\bar{v}}^{\,b}\,\delta(v,\bar{v})=(-1)^{a+b}\,\partial_v^{\,a} \partial_{\bar{v}}^{\,b}\, f(v,\bar{v})\at{7}{1.8}{v=0,\bar{v}=0}\,,
\end{align}
for any pair of non-negative integers $a,b\in\mathbb{N}_0$. Monomials of complex variables integrate to derivatives of the complex delta distribution according to
\begin{align}\label{eq:complex_integral}
    \int_{\mathbb{C}}\!\d w\d\bar{w}\ \bar{w}^a w^b \e{i\kappa(v\bar{w}+\bar{v}w)}=\left(-\frac{i}{\kappa}\right)^{a+b+2}\partial_{v}^{\,a}\partial_{\bar{v}}^{\,b}\,\delta(v,\bar{v})\,,
\end{align}
for any $\kappa\in\mathbb{R}$ and $v\in\mathbb{C}$; it is then straightforward to determine the normalisation of the Hermitian product, as well as the orthogonality relation for monomials of the oscillator variables
\begin{align}
    \left(\left(v^{(1)}_m\right)^{a}\left(v^{(2)}_m\right)^{b}\,,\left(v^{(1)}_m\right)^{c}\left(v^{(2)}_m\right)^{d}\right)=\frac{a!b!}{(4m)^{a+b}}\delta_{a,d}\delta_{b,c}\,.
\end{align}

\subsection{Generators}
To find the oscillator construction of $\bms_3$ we take care of the normal ordering in the $\bms_3$ generators \eqref{LMbmsa}; plug in the differential representations of the oscillators, as given in equations \eqref{adiff} and \eqref{ahatdiff}; use the definitions \eqref{a0lambdamu}, and then simplify. Denoting the generators in the oscillator construction by lowercase, we arrive at
\begin{subequations}
\begin{align}
\label{eq:l0_app}
    l_0 &= \Delta+\sum_{n=1}^\infty n \del{v^{(1)}_n \partial_{v^{(1)}_n} + v^{(2)}_n \partial_{v^{(2)}_n}} \,,\\
\begin{split}
        l_k &= \sum_{n=1}^\infty n\del{ {v^{(1)}_n} \partial_{ v^{(1)}_{k+n}} + v^{(2)}_n \partial_{ v^{(2)}_{k+n}} } 
        -\frac18 \sum_{n=1}^{k-1} \del{ \partial_{v^{(1)}_n} \partial_{ v^{(2)}_{k-n}} + \partial_{v^{(2)}_n} \partial_{ v^{(1)}_{k-n}} } \\
        &\quad +\frac{i}{2} \left[ (\lambda_1 - \lambda_2) - i k(\mu_1 - \mu_2)  \right] \partial_{ v^{(1)}_k } +\frac{i}{2} \left[(\lambda_1 + \lambda_2) - i k(\mu_1 + \mu_2) \right] \partial_{ v^{(2)}_k } \,,
\end{split}\\
\begin{split}
        l_{-k} &= \sum_{n=1}^\infty (k+n)\del{ {v^{(1)}_{k+n} } \partial_{ v^{(1)}_{n}} + v^{(2)}_{k+n} \partial_{ v^{(2)}_{n}} } 
        -2 \sum_{n=1}^{k-1} n(k-n)\del{ {v^{(1)}_n} { v^{(2)}_{k-n}} + {v^{(2)}_n} { v^{(1)}_{k-n}} } \\
        &\quad - i 2 k\left[ (\lambda_1 + \lambda_2) + i k(\mu_1 + \mu_2)  \right] { v^{(1)}_k } -  i2 k\left[(\lambda_1 - \lambda_2) +i  k(\mu_1 - \mu_2) \right] { v^{(2)}_k } \,,
\end{split}
\end{align}
\end{subequations}
as well as 
\begin{subequations}
\begin{align}
\label{eq:m0_app}
    m_0 &= \xi+\sum_{n=1}^\infty n v^{(1)}_n \partial_{ v^{(2)}_n } \,, \\
    m_k &= \sum_{n=1}^\infty n v^{(1)}_n \partial_{ v^{(2)}_{k+n} } - \frac18 \sum_{n=1}^{k-1} \partial_{ v^{(2)}_{k-n} } \partial_{ v^{(2)}_{n} } + \frac{i}{2}\left[( \lambda_1 - \lambda_2) - ik(\mu_1 - \mu_2) \right]\partial_{v^{(2)}_k} \,, \\
    m_{-k} &= \sum_{n=1}^\infty (k+n) v^{(1)}_{k+n} \partial_{ v^{(2)}_{n} } - 2 \sum_{n=1}^{k-1} n(k-n){ v^{(1)}_{k-n} } { v^{(1)}_{n} } - i2k\left[( \lambda_1 - \lambda_2) + ik(\mu_1 - \mu_2) \right]{v^{(1)}_k} \,.
\end{align}
\end{subequations}
In \eqref{eq:l0_app} and \eqref{eq:m0_app} we have identified the scaling dimension $\Delta$ and rapidity $\xi$ as
\begin{subequations}\label{defDeltaXi}
\begin{align}
    \label{eq:def_Delta}
    \Delta &\equiv  \lambda_1^2 - \lambda_2^2 + \mu_1^2  - \mu_2^2\,,\\
    \label{eq:def_Xi}
    \xi &\equiv  \frac12\left[\del{ \lambda_1 - \lambda_2 }^2 + (\mu_1 - \mu_2)^2\right]\,.
\end{align}
\end{subequations}
Furthermore, the above generators satisfy the $\bms_3$ algebra with central charges $c_{\text{\tiny L}} = 2 + 24\del{\mu_1^2 - \mu_2^2}$ and $c_{\text{\tiny M}} = 12\del{\mu_1 - \mu_2}^2$.

Finally, we will make use of the abbreviations 
\begin{subequations}\label{eq:def_AkBk}
\begin{align}
    A_k&=\frac{i}{2}\,\left[\lambda_1-\lambda_2-i k(\mu_1-\mu_2)\right]\,,   &   B_k&=\frac{i}{2}\,\left[\lambda_1+\lambda_2-i k(\mu_1+\mu_2)\right]\,, \\ 
    \hat{A}_k&=-\frac{i}{2}\,\left[\lambda_1-\lambda_2+i k(\mu_1-\mu_2)\right]\,,  &   \hat{B}_k&=-\frac{i}{2}\,\left[\lambda_1+\lambda_2+i k(\mu_1+\mu_2)\right]\,.
\end{align}
\end{subequations}
Using the definitions \eqref{defDeltaXi} together with the values of the central charges, the coefficients defined above may be expressed in terms of familiar quantities as
\begin{subequations}
\begin{align}
    A_k&=\mp\frac{i}{2}\sqrt{2\xi-\frac{c_{\text{\tiny M}}}{12}}\pm k\sqrt{\frac{c_{\text{\tiny M}}}{48}}\,,
    &B_k&=\pm i\frac{c_{\text{\tiny L}}-2-24\Delta}{48\sqrt{2\xi-\frac{c_{\text{\tiny M}}}{12}}}\pm  k\frac{c_{\text{\tiny L}}-2}{\sqrt{48\frac{c_{\text{\tiny M}}}{12}}}\,,\\
    \hat{A}_k&=\pm\frac{i}{2} \sqrt{2\xi-\frac{c_{\text{\tiny M}}}{12}}\pm  k\sqrt{\frac{c_{\text{\tiny M}}}{48}}\,,
    &\hat{B}_k&=\mp i \frac{c_{\text{\tiny L}}-2-24\Delta}{48\sqrt{2\xi-\frac{c_{\text{\tiny M}}}{12}}}\pm  k\frac{c_{\text{\tiny L}}-2}{48\sqrt{\frac{c_{\text{\tiny M}}}{12}}}\,,
\end{align}
\end{subequations}
where the sign choices for the $k$-dependent terms are correlated with each other, and the sign choices for the $k$-independent terms are correlated with each other.

%%%%%%%%%%%%%%%%%%%%%%%%%%%%%%%%%%%%%%%%%%%%%%%%%%%%%%%%%%%
%%%%%%%%%%%%%%%%%%%%%%%%%%%%%%%%%%%%%%%%%%%%%%%%%%%%%%%%%%%
%%%%%%%%%%%%%%%%%%%%%%%%%%%%%%%%%%%%%%%%%%%%%%%%%%%%%%%%%%%

\section{Proof of Unique Solution}\label{app:uniqueness}
In this section we prove that the level-two $\bms_3$ wave functions admit unique solutions; we show it for finite central charge $c_{\text{\tiny M}}$ and comment on the semi-classical limit $c_{\text{\tiny M}}\to\infty$. The discussion for the semi-classical limit may be considered to be part of the proof of exponentiation presented in subsection \ref{sec:exponentiation}. Parts of the analysis below follow methods presented in \cite{Besken:2019jyw}.

\subsection{Proof of Unique Solution for \texorpdfstring{$F(\eta,\nu)$}{Fetanu}}\label{sec:uniF}
We focus on $\psi_{\Delta_{1,2},\xi_{1,2};\Delta,\xi}(t,x,0,0;v)$, similar arguments apply to the dual wave function $\chi_{\Delta_{1,2},\xi_{1,2};\Delta,\xi}(t,x,0,0;\bar{v}).$ To get a valid wave function for the point configuration under consideration, $F(\eta, \nu)$ in the ansatz \eqref{wfsolnk0BMS} must satisfy the differential equations \eqref{diffeqpsilvl2BMS} for $n\geq 1.$\footnote{If we consider a general point configuration with $t_2$ and $x_2$ being non-zero, the $n=-1$ equation may be used to reinstate the dependence on these coordinates.} In order to express the differential equations in terms of $F(\eta,\nu)$ it is beneficial to write the $\bms_3$ generators $l_0, m_0$ and $l_k, m_k$, with $k\geq1$, using the variables $\eta_m$ and $\nu_m$; setting $t=1$ and $x=0$ for simplicity, we get
\begin{subequations}
\begin{align}
    l_0 &= \Delta + \sum_{n=1}^\infty n \left(\eta_n \partial_{\eta_n} + \nu_n \partial_{\nu_n} \right) \, ,  \\
    l_k &= \sum_{n=1}^\infty n \left( \eta_n \partial_{\eta_{k+n}} + \nu_n \partial_{ \nu_{k+n}} \right) 
        -\frac{1}{4} \sum_{n=1}^{k-1} \partial_{\eta_n} \partial_{ \nu_{k-n}} + A_k \partial_{ \eta_k } + B_k \partial_{ \nu_k } \, , \\
    m_0 &= \xi + \sum_{n=1}^\infty n \, \eta_n \partial_{\nu_n} \, , \\ \label{eq:lkapp}
    m_k &= \sum_{n=1}^\infty n \eta_n \partial_{ \nu_{k+n} } - \frac18 \sum_{n=1}^{k-1} \partial_{ \nu_{k-n} } \partial_{ \nu_{n} } + A_k \partial_{\nu_k} \, .
\end{align}
\end{subequations}
Moreover, to implement the action of $\mathcal{L}^{(\Delta_1,\xi_1)}_n$ and $\mathcal{M}^{(\Delta_1,\xi_1)}_n$ we make use of the identities 
\begin{subequations}
\begin{align}
    \partial_x F(\eta,\nu)\Big|_{t=1,x=0} &= \sum\limits_{m=1}^\infty m \, \eta_m \, \partial_{\nu_m} F(\eta,\nu)\,, \\
    \partial_t F(\eta,\nu)\Big|_{t=1,x=0} &= \sum\limits_{m=1}^\infty m \left( \eta_m \, \partial_{\eta_m} +  \nu_m \, \partial_{\nu_m}  \right)F(\eta,\nu) \, ,
\end{align}
\end{subequations}
which result from applying the chain rule to $F(\eta,\nu)$. Taking all of the above quantities together, the differential equations \eqref{diffeqpsilvl2BMS} with $n \geq 1$ give rise to the equations
\begin{subequations}\label{eq:lk_mk_F}
    \begin{align}
    l_k F(\eta,\nu) &=(l_0 + k \Delta_1 - \Delta_2) F(\eta,\nu) \, , \label{eq:lkFapp} \\ 
    m_k F(\eta,\nu) &=(m_0 + k \xi_1 - \xi_2) F(\eta,\nu) \, . \label{eq:mkFapp}
\end{align}
\end{subequations}

We now assume that $F(\eta,\nu)$ takes the form of a power series expansion in the variables $\eta_m$ and $\nu_m$,
\begin{equation}
    F(\eta,\nu) = \sum\limits_{n=0}^\infty F_n\,,
\end{equation}
where each $F_n$ is a linear combination of every possible linearly independent monomial of the form $\eta_{j_1} \eta_{j_2}\cdots \eta_{j_p} \nu_{k_1} \nu_{k_2} \cdots \nu_{k_q}$, with the ordering of variables such that $j_1 \leq j_2 \leq \cdots \leq j_p$ and $k_1 \leq k_2 \leq \cdots \leq k_q$, and the sum of the indices satisfies $\sum_{i=1}^p j_i + \sum_{i=1}^q k_i = n$. Hence, the number of terms in $F_n$ is given by  
\begin{equation}
p_2(n) = \sum\limits_{k=0}^n p(k) p(n-k)\,, 
\end{equation}
where $p(k)$ is the number of partitions of the number $k$. We choose the normalisation $F_0=1$.

We may view $F_n$ as an element of a vector space $V_n$ which is spanned by the monomials $\eta_{j_1} \eta_{j_2}\cdots \eta_{j_p} \nu_{k_1} \nu_{k_2} \cdots \nu_{k_q}$. The dimension of the vector space $V_n$ is $p_2(n)$, i.e. equal to the number of basis monomials. Due to the action of the derivative in the differential operators $l_k$ and $m_k$ we note that $l_k (V_n) \subseteq V_{n-k}$ as well as $m_k (V_n) \subseteq V_{n-k}$ for $k \leq n.$

\subsubsection{Building a Linear Set of Equations}

To specify $F_n$, and in turn $F(\eta,\nu)$, we have to know the coefficients in front of the $p_2(n)$ basis vectors of $V_n$. Our strategy is to show that there are as many coefficients as equations for these coefficients, and that the solution to this set of linear equations is unique. 

In terms of $F_n$ the equations \eqref{eq:lk_mk_F} read, for $k\leq n$,
\begin{subequations}
\begin{align}
    l_k F_n &= l_0 F_{n-k} + (k\Delta_1 - \Delta_2) F_{n-k} \, ,\label{eq:lkFn} \\
    m_k F_n &= m_0 F_{n-k} + (k\xi_1 - \xi_2) F_{n-k} \, ,
\end{align}
\end{subequations}
where we have used that $l_0 (V_{m}) \subseteq V_{m}$ and $m_0 (V_m) \subseteq V_m$ for $m \in \mathbb{N}_0$, and the right-hand side contains $F_{n-k}$ because the two sides must belong to the same vector space $V_{n-k}$. The form of $l_0$ gives rise to the eigenvalue equation $l_0 v = \del{\Delta + m}v$ for any element $v \in V_m$; this observation allows us to simplify \eqref{eq:lkFn} further, arriving at
\begin{subequations}\label{lk_mk_recursive}
\begin{align}
       l_k F_n &= \beta_{k,n} \, F_{n-k} \, ,\label{eq:lkFnbeta} \\
    m_k F_n &= m_0 F_{n-k} + (k\xi_1 - \xi_2) F_{n-k} \, ,
\end{align}
\end{subequations}
with $\beta_{k,n}\equiv \Delta + n - k +k\Delta_1 - \Delta_2$. Unfortunately, we cannot constrain $m_0 F_{n-k}$ further since $m_0$ does not give rise to a simple action on the vector space $V_m$.

The set of equations \eqref{lk_mk_recursive} displays some recursive traits which we will use to our advantage; remembering that $F_0=1$, we may reduce the index of $F_n$ until the right-hand side only contains numbers. To this end, we evaluate a string of operators $l_{k_q} \cdots l_{k_2} l_{k_1}$ acting on $F_n$ by applying equation \eqref{eq:lkFn} recursively; we obtain
\begin{equation}\label{eq:stringlF}
    l_{k_q} \cdots l_{k_1} F_n = \beta_{k_q, n-K+k_q} \cdots \beta_{k_2,n-k_1} \, \beta_{k_1,n} \, F_{n- K}\,,
\end{equation}
provided that $K \equiv \sum_{i=1}^q k_i$ is less than $n$. Since $m_0$ does not have a simple eigenvalue equation, we cannot evaluate a string of operators $m_{j_p} \cdots m_{j_2} m_{j_1}$ acting on $F_{n-K}$ unless $J\equiv \sum_{i=1}^p j_i$ is equal to $n-K$; in this case we find
\begin{align}\label{eq:stringmF1}
m_{j_p} \cdots m_{j_2} m_{j_1} F_{n-K} &= (m_0 + j_p \xi_1 - \xi_1) \cdots (m_0 + j_2 \xi_1 - \xi_1) (m_0 + j_1 \xi_1 - \xi_1) F_0\,,
\end{align}
where we used that the generators $m_{j_i}$ mutually commute so that each product of $m_{j_i}$ acts directly on $F_\kappa$, for the suitable level $\kappa$. We may apply $m_0 F_0 = \xi$ recursively to equation \eqref{eq:stringmF1}, which yields
\begin{align}\label{eq:stringmF}
m_{j_p} \cdots m_{j_2} m_{j_1} F_{n-K} &= \gamma_{j_p} \cdots \gamma_{j_2} \gamma_{j_1}\,,
\end{align}
where $\gamma_j = \xi + j \xi_1 - \xi_1$, for the case $J =n-K$. The two equations \eqref{eq:stringlF} and \eqref{eq:stringmF} bring us to the general form of repeated action of generators to be\footnote{Note that the ordering of $m_{j_p} \cdots m_{j_2} m_{j_1}  l_{k_q} \cdots l_{k_1}$ is crucial in equation \eqref{eq:mixedstringF}. In particular, we are not able to easily evaluate the reversed ordering of $l$ and $m$ operators, i.e. $l_{k_q} \cdots l_{k_1}m_{j_p} \cdots m_{j_2} m_{j_1}$.}
\begin{equation}\label{eq:mixedstringF}
m_{j_p} \cdots m_{j_2} m_{j_1}  l_{k_q} \cdots l_{k_1} F_n = \gamma_{j_p} \cdots \gamma_{j_2} \gamma_{j_1} \beta_{k_q, n-K+k_q} \cdots \beta_{k_2,n-k_1} \, \beta_{k_1,n}
\end{equation}
in the case of $\sum_{i=1}^p j_i + \sum_{i=1}^q k_i = n.$

\subsubsection{Existence and Uniqueness of the Solution}
The left-hand side of equation \eqref{eq:mixedstringF} is a linear combination of the $p_2(n)$ unknown coefficients of the basis monomials. There are $p_2(n)$ possible strings of ordered operators $m_{j_p} \cdots m_{j_2} m_{j_1}  l_{k_q} \cdots l_{k_1}$ with  $J + K = n$. Hence, by considering all actions of the form $m_{j_p} \cdots m_{j_2} m_{j_1}  l_{k_q} \cdots l_{k_1} F_n$ we form a linear system of $p_2(n)$ equations with $p_2(n)$ unknowns. To prove that this system is solvable it is sufficient to show that the matrix $\mathscr{M}_n$, whose entries are given by $m_{j_1} \cdots m_{j_p}  l_{k_1} \cdots l_{k_q}(\eta_{j'_1} \cdots \eta_{j'_{r}} \nu_{k'_1}  \cdots \nu_{k'_{s}})$ and which characterises the system of equations, is invertible except for at most a discrete set of exceptional cases; we will do this by arguing that the determinant of $\mathscr{M}_n$ is generically non-vanishing.

We will continue in a two-step procedure. In the first step we simplify the task by replacing the operators $l_k$ by ${l}^\circ_k = A_k\partial_{\eta_k}+B_k\partial_{\nu_k}$ and $m_k$ by ${m}^\circ_k = A_k\partial_{\nu_k}$. It is always possible to find a basis such that the simplified matrix $\mathscr{M}^\circ_n$ has an upper triangular form, with products of $A_k$ on its diagonal; we postpone the proof of this statement to the next subsection. The determinant of $\mathscr{M}^\circ_n$ is the product of its diagonal elements, i.e. products of $A_k$. Using that $A_k$ is given by \eqref{Ak_Bk}, the determinant vanishes only if  
\begin{equation}
    \xi = \frac{c_{\text{\tiny M}}}{24}\left(1-k^2\right)
    \label{eq:xi_k}
\end{equation}
with $k\in \mathbb{N}$; this means that there is only a discrete set of values of $\xi$ for which the determinant of $\mathscr{M}^\circ_n$ vanishes.

As the second step we need to consider the matrix $\mathscr{M}_n$ corresponding to the equations using the full operators $l_k$ and $m_k$. A product of the operators $l_k$, $m_k$ can be re-expressed in terms of sums of ${l}^\circ_k$, ${m}^\circ_k$ and $l_k - {l}^\circ_k$, $m_k - {m}^\circ_k$; where the differences between the full and simplified operators are independent of $\xi$ and $c_{\text{\tiny M}}$. Hence, the determinant of $\mathscr{M}_n$ and the determinant of $\mathscr{M}^\circ_n$ have features in common: both are polynomials of the same degree in $y \equiv \sqrt{2\xi- c_{\text{\tiny M}}/12}$, and they have the same  coefficient of the largest power of $y$. Consequently, the determinant of $\mathscr{M}_n$ may vanish only for a discrete set of values $y$, generically without relation to \eqref{eq:xi_k}. We have thus shown that the system of equations for the coefficients of $F_n$ admits a unique solution; this means that a solution for $F(\eta,\nu)$ is unique, too.

\subsubsection{Proof of Unique Solution for \texorpdfstring{$S(\sigma, \kappa)$}{S}}\label{sec:uniqS}
In the semi-classical limit we make the ansatz $F(\sigma,\kappa) = \exp(\mu^2 S(\sigma,\kappa))$. The proof of a unique solution for  $S(\sigma, \kappa)$ goes along the same lines as above. However, there are two differences which we would like to highlight. First, we treat the equations in terms of $S(\sigma, \kappa)$, not in terms of $F(\sigma, \kappa)$. Hence, the expansion of $S(\sigma, \kappa)$ in terms of its variables is reorganised compared to the expansion of $F(\sigma, \kappa)$. Second, due to the semi-classical limit we may drop the second derivative terms in $l_k$ and $m_k$. Neither of the changes mentioned in this subsection invalidate the arguments above concerning the uniqueness of the wave function.

\subsection{Upper-triangularity of \texorpdfstring{$\mathscr{M}^\circ_n$}{M}}
We now analyse the evaluation of the string of operators ${m}^\circ_{j_1} \cdots {m}^\circ_{j_p}  {l}^\circ_{k_1} \cdots {l}^\circ_{k_q}$ on monomials of the form $\eta_{j'_1} \cdots \eta_{j'_{r}} \nu_{k'_1}  \cdots \nu_{k'_{s}}$. Both sets of indices are ordered and add up to $n$, i.e. both sets are partitions of $n$. We will prove that there exists an ordering of the partitions of the index variables of the operators and monomials such that the matrix $\mathscr{M}^\circ_n$, with entries ${m}^\circ_{j_1} \cdots {m}^\circ_{j_p}  {l}^\circ_{k_1} \cdots {l}^\circ_{k_q}(\eta_{j'_1} \cdots \eta_{j'_{r}} \nu_{k'_1}  \cdots \nu_{k'_{s}})$, is upper triangular. 

We will prove our claim by induction; we start with the case $n=1$. In this case the matrix reads
\begin{align*}
    \mathscr{M}^\circ_1=\begin{pmatrix}l^\circ_1\eta_1 & l^\circ_1\nu_1\\ m^\circ_1\eta_1 & m^\circ_1 \nu_1\end{pmatrix}\,,
\end{align*}
which illustrates our choice of basis $(l^\circ_1,m^\circ_1)$ and $(\eta_1,\nu_1)$, where partitions of generator indices are associated to rows, whereas partitions of oscillator indices are associated to columns of the respective matrix $\mathscr{M}^\circ_n$. Since $m^\circ_1 \eta_1 =0$  we find that $\mathscr{M}^\circ_1$ is upper triangular with $A_1$ on the diagonal.

Assuming the matrix  $\mathscr{M}^\circ_n$ to be upper triangular and having products of $A_k$ as entries on its diagonal, our task is to show that $\mathscr{M}^\circ_{n+1}$ has the same properties. The matrix entries ${m}^\circ_{j_1} \cdots {m}^\circ_{j_p}  {l}^\circ_{k_1} \cdots {l}^\circ_{k_q}(\eta_{j'_1} \cdots \eta_{j'_{r}} \nu_{k'_1}  \cdots \nu_{k'_{s}})$ of $\mathscr{M}^\circ_{n+1}$ can be non-zero only if the sets $\{j_1, \dots j_p, k_1, \dots k_q\}$ and $\{j'_1, \dots j'_r, k'_1, \dots k'_s\}$ belong to the same partition of $n+1$; this is evident from the action of the differential operators ${l}^\circ_{k}$ and ${m}^\circ_k$ on the basis monomials. Hence, the matrix $\mathscr{M}^\circ_{n+1}$ is block diagonal: each block of the matrix $\mathscr{M}^\circ_{n+1}$ corresponds to a particular partition of $n+1$. 

The next step is to prove that for a given partition of $n+1$ its corresponding block of $\mathscr{M}^\circ_{n+1}$ is upper-triangular. Note that any partition of the number $n+1$, apart from the trivial partition just containing $n+1$ times the number one, may be obtained from a partition of the number $n$ by raising one of the numbers by one without violating the (decreasing) ordering of the numbers in the partition. For example, the partition $(4,2)$ of the number $6$ can be obtained from the partition $(4,1)$ of the number $5$. By the induction assumption, the matrix $\mathscr{M}^\circ_n$ is upper triangular and hence also the block corresponding to the partition $n$. Thus, keeping the ordering of the operators and monomials (but raising one of the indices to match the partition of $n+1$) the block of the matrix $\mathscr{M}^\circ_{n+1}$ corresponding to that partition of $n+1$ is also upper triangular. By this construction, the diagonal of  $\mathscr{M}^\circ_{n+1}$ contains products of $A_k$ only. 

We are left with the trivial partition $1+\dots +1$ of $n+1.$ In this case we have to find a suitable ordering of the operators $({m}^\circ_1)^p ({l}^\circ_1)^q$ with $p+q=n+1$ and the monomials $(\eta_1)^r (\nu_1)^s$ with $r+s=n+1.$  Since ${m}^\circ_1$ is a derivative with respect to $\nu_1$ we have
\begin{equation}
({m}^\circ_1)^p ({l}^\circ_1)^q \left( (\eta_1)^r (\nu_1)^s \right) = 0 \, \qquad \textrm{for} \quad p>s \, .
\end{equation}
Hence ordering the operators $({m}^\circ_1)^p ({l}^\circ_1)^{n+1-p}$ with increasing $p=0,\dots, n+1$  and the monomials $(\eta_1)^{n+1-r} (\nu_1)^{r}$ with increasing $r=0, \dots, n+1$ we obtain an upper triangular matrix with powers of $A_1$ on its diagonal.

This completes the proof by induction. Hence we conclude that the matrix  $\mathscr{M}^\circ_n$ has an upper-triangular form for a specific ordering of the operators and monomials, with products of $A_k$ on its diagonal.

%%%%%%%%%%%%%%%%%%%%%%%%%%%%%%%%%%%%%%%%%%%%%%%%%%%%%%%%%%%
%%%%%%%%%%%%%%%%%%%%%%%%%%%%%%%%%%%%%%%%%%%%%%%%%%%%%%%%%%%
%%%%%%%%%%%%%%%%%%%%%%%%%%%%%%%%%%%%%%%%%%%%%%%%%%%%%%%%%%%

\section{Determining the Level-one Wave Function}\label{sec:lvl1soln}
In this appendix we solve the differential equations \eqref{psilvl1BMS} for the level-one wave function. The result for the dual wave function follows from applying the same procedure detailed below to the differential equations \eqref{chilvl1BMS}. 

We first turn our attention to equation \eqref{psilvl1BMSa} for $n=0$, which reads
\begin{equation}
\left(l^{(\Delta,\xi)}_0 + \mathcal{L}^{(\Delta_2,\xi_2)}_0\right)\psi_{\Delta_2,\xi_2;\Delta,\xi}(t_2,x_2;v)=0\,.
\end{equation}
Plugging in the generators \eqref{eq:l0bms} and \eqref{curlyL} the above equation reads
\begin{equation}\label{eq:appCloeq}
    -t_2\partial_{t_2}\Psi-x_2\partial_{x_2} \Psi+\sum_{n=1}^{\infty}n\left(v_{n}^{(1)}\partial_{v_{n}^{(1)}}+v_{n}^{(2)}\partial_{v_{n}^{(2)}}\right)\Psi+\left(\Delta-\Delta_2\right)\Psi=0\,,
\end{equation}
where we have abbreviated $\Psi\equiv\psi_{\Delta_2,\xi_2;\Delta,\xi}(t_2,x_2;v)$ for simplicity. The above differential equation may be solved by applying the method of characteristics. The Lagrange-Charpit equations corresponding to \eqref{eq:appCloeq} are
\begin{equation}
    -\frac{\dd t_2}{t_2}=-\frac{\dd x_2}{x_2}=\sum\limits_{n=1}^\infty  \frac{\dd v_n^{(1)}}{n \, v_n^{(1)}}=\sum\limits_{n=1}^\infty \frac{\dd v_n^{(2)}}{n \, v_n^{(2)}}=\frac{\dd\Psi}{\left(\Delta-\Delta_2\right)\Psi}\,.
\end{equation}
The characteristic curves $c_i$ are given by
\begin{subequations}
\begin{align}
        c_1&=\frac{x_2}{t_2}\,,\,&c_2&=\sum\limits_{n=1}^\infty v_n^{(1)}t_2^n\,,\\
        c_3&=\sum\limits_{n=1}^\infty v_n^{(2)}t_2^n\,,\,&c_4&=\Psi t_2^{\Delta-\Delta_2}\,;
\end{align}
\end{subequations}
and using the method of characteristics we find the solution of the differential equation \eqref{eq:appCloeq} to be
\begin{equation}\label{psif1}
    \psi_{\Delta_2,\xi_2;\Delta,\xi}(t_2,x_2;v)=t_2^{\Delta-\Delta_2}f_1\left(\frac{x_2}{t_2}\,,\,\sum_{n=1}^{\infty}v_n^{(1)}t_2^n\,,\,\sum_{n=1}^{\infty}v_n^{(2)}t_2^n\right)\,,
\end{equation}
where $f_1(\,\cdots)$ is an unknown function to be determined, and we have returned to the original notation of the wave function.

It follows from the definition of the primary function given in \eqref{BMSprimary} that we may impose the boundary condition $\lim_{t_2\rightarrow 0}\psi_{\Delta_2,\xi_2;\Delta,\xi}(t_2,0;v)=\mathbf{1}$. This boundary condition implies that the unknown function in \eqref{psif1} is an exponential function, and the prefactor $t^{\Delta-\Delta_2}$ enforces that $\Delta=\Delta_2$; hence the wave function is of the form
\begin{equation}
    \psi_{\xi_2;\Delta,\xi}(t_2,x_2;v)=\exp[f\left(\frac{x_2}{t_2}\,,\,\sum_{n=1}^{\infty}v_n^{(1)}t_2^n\,,\,\sum_{n=1}^{\infty}v_n^{(2)}t_2^n\right)]\,.
\end{equation}

Inserting the above partial solution into equation \eqref{psilvl1BMSb} for $n=0$ and $n=1$, leads to two differential equations, which when combined become
\begin{equation}
    \partial_{c}f\left(a,b,c\right)=\frac{\xi+\xi_2}{A_1}\,;
\end{equation}
where we have relabeled the arguments as $a=\frac{x_2}{t_2}$, $b=\sum_{n=1}^\infty v_n^{(1)}t_2^n,$ as well as $c= \sum_{n=1}^{\infty} v_n^{(2)} t_2^n$, and $A_1$ is defined by (\ref{Ak_Bkfull}). Integrating the above equation and reinserting the solution into the wave function, we arrive at
\begin{equation}\label{psig}
    \psi_{\xi_2;\Delta,\xi}(t_2,x_2;v)=\exp[\frac{\xi+\xi_2}{A_1}\sum_{n=1}^\infty v_n^{(2)}t_2^n + g\left(\frac{x_2}{t_2}\,,\,\sum_{n=1}^\infty v_n^{(1)}t_2^n\right)]\,,
\end{equation}
where $g\left(\,\cdots\right)$ is an unknown function which arises as a constant of integration. Inserting the partial solution \eqref{psig} back into equation \eqref{psilvl1BMSb} with $n=0$ results in the differential equation
\begin{equation}
    \partial_a g(a,b)=\xi-\xi_2 + \frac{\xi+\xi_2}{A_1} t_2\partial_{t_2}b\,,
\end{equation}
where $a=\frac{x_2}{t_2}$ and $b=\sum_{n=1}^{\infty}v_n^{(1)}t_2^n$. The above equation can be integrated to obtain
\begin{equation}
    g\left(\frac{x_2}{t_2}\,,\,\sum_{n=1}^\infty v_n^{(1)}t_2^n\right)=\left( \xi-\xi_2+\frac{\xi+\xi_2}{A_1} \sum_{n=1}^\infty n t_2^{n}v_n^{(1)} \right) \frac{x_2}{t_2}+h\left(\sum_{n=1}^\infty v_n^{(1)}t_2^n\right)\,,
\end{equation}
where $h\left(\,\cdots\right)$ is again an unknown function arising as a constant of integration. With the above expression the wave function takes the form
\begin{equation}
    \psi_{\xi_2;\Delta,\xi}(t_2,x_2;v)=\exp[\frac{\xi+\xi_2}{A_1}\sum\limits_{n=1}^\infty\left(v_n^{(2)}t_2^n+n x_2 t_2^{n-1} v_n^{(1)}\right)+\frac{x_2}{t_2}\left(\xi-\xi_2\right)+h\left(\sum\limits _{n=1}^\infty v_n^{(1)}t_2^n\right)]\,.
\end{equation}
Plugging the partial solution above into equation \eqref{psilvl1BMSb} with $n=1$ returns the relation $\xi+\xi_2=2\xi_2$,
which requires that $\xi_2=\xi$; thus the wave function can be further simplified into
\begin{equation}
    \psi_{\Delta,\xi}(t_2,x_2;v)=\exp[\frac{2\xi}{A_1}\left( \sum\limits _{n=1}^\infty v_n^{(2)}t_2^n+\frac{x_2}{t_2} \sum\limits _{n=1}^\infty n t_2^{n}v_n^{(1)}\right)+h\left(\sum_{n=1}^\infty v_n^{(1)}t_2^n\right)]\,.
\end{equation}
Finally, inserting this partial wave function solution into equation \eqref{psilvl1BMSa} for $n=1$leads to the differential equation 
\begin{equation}
    \partial_{b}h\left(b\right)=4\hat{B}_1\,,
\end{equation}
with $b=\sum_{n=1}^\infty v_n^{(1)}t_2^n$ and $\hat{B}_1$ is defined by (\ref{Ak_Bkfull}). The solution of this differential equation is
\begin{equation}
h\left(\sum\limits_{n=1}^{\infty}v_n^{(1)}t_2^n\right)=4\hat{B}_1 \sum\limits_{n=1}^{\infty}v_n^{(1)}t_2^n\,.
\end{equation} 

Combining the results, the level-one wave function is thus fully determined up to an overall constant to be
\begin{equation}\label{eq:sol_lvl1_wave}
    \psi_{\Delta,\xi}(t_2,x_2;v) = \exp\left[ 4 \hat{A}_1\sum_{n=1}^\infty \del{ t_2^n v^{(2)}_n + n\,x_2\,t_2^{n-1} v^{(1)}_n} + 4\hat{B}_1  \sum_{n=1}^\infty t_2^{n} v^{(1)}_n\right]\,,
\end{equation}
where we have used the relations
\begin{subequations}
\begin{align}
    2\Delta & = 4\left(A_1\hat{B}_1+\hat{A}_1B_1\right)\,,\\
    2\xi & = 4A_1\hat{A}_1\,.
\end{align}
\end{subequations}

Note that we did not use all of the differential equations \eqref{psilvl1BMS} when solving for the level-one wave function. We explicitly checked that the solution \eqref{eq:sol_lvl1_wave} satisfies the differential equations \eqref{psilvl1BMS} with $n=-1$ as well as $n= 2$, which means that the full set is satisfied.

%%%%%%%%%%%%%%%%%%%%%%%%%%%%%%%%%%%%%%%%%%%%%%%%%%%%%%%%%%%
%%%%%%%%%%%%%%%%%%%%%%%%%%%%%%%%%%%%%%%%%%%%%%%%%%%%%%%%%%%
%%%%%%%%%%%%%%%%%%%%%%%%%%%%%%%%%%%%%%%%%%%%%%%%%%%%%%%%%%%

\section{Solutions for Generating Functions}\label{apdx:differentialequations}
In this section we present the methodologies that were used to solve the two differential equations
\begin{align}
         \partial_{\tau}\left(\tau\cdot C(\tau)\right)&=-\sqrt{\frac{3}{\tilde{c}_{\text{\tiny M}}}} C(\tau)D(\tau)-\sqrt{\frac{48}{\tilde{c}_{\text{\tiny M}}}}\frac{\tilde{\Delta}_3 \tau^2}{(1-\tau)^2}-\frac{\tilde{c}_{\text{\tiny L}}}{2\tilde{c}_{\text{\tiny M}}}\partial_{\tau}\left(\tau\cdot D(\tau)\right)\,, \label{eq:C_diff}\\
        \partial_{\tau}\left(\tau\cdot D(\tau)\right)&=-\frac{1}{8}\sqrt{\frac{48}{\tilde{c}_{\text{\tiny M}}}}\left(D(\tau)\right)^2-\sqrt{\frac{48}{\tilde{c}_{\text{\tiny M}}}}\frac{\tilde{\xi}_3 \tau^2}{(1-\tau)^2} \, , \label{D_diff1} 
\end{align}
which appeared when determining the heavy-light vacuum $\bms_3$-block in subsection \ref{subsec:HeavyLight} of the main text. We will solve equation \eqref{D_diff1} for $D(\tau)$ and use that solution in \eqref{eq:C_diff} to find $C(\tau)$.

%%%%%%%%%%%%%%%%%%%%%%%%%%%%%%%%
%%%%%%%%%%%%%%%%%%%%%%%%%%%%%%%%%

\subsection{Solving \texorpdfstring{$D(\tau)$}{D(t)}}
Equation \eqref{D_diff1} is known as a Riccati equation and may be solved by first finding a particular solution $D_{\text{\tiny p}}(\tau)$ which in turn is used to find the general solution via
\begin{equation}\label{Dgeneralexp}
    D(\tau) = D_{\text{\tiny p}}(\tau) + y(\tau)\,,
\end{equation}
where $y(\tau)$ must be determined in a second step. For the equation at hand a particular solution can be found by making the ansatz $D(\tau) = \frac{a \tau + b}{1-\tau}$ and solving the resulting equation for $a$ and $b$; the solution then reads
\begin{equation}
    D_{\text{\tiny p}}(\tau)=\sqrt{\frac{\tilde{c}_{\text{\tiny M}}}{48}}\frac{4}{1-\tau}\left(\tau\left(1+\sqrt{1-\frac{24\tilde{\xi}_3}{\tilde{c}_{\text{\tiny M}}}}\right)-2\right)\,.
\end{equation}
Using the above expression in \eqref{Dgeneralexp} gives a first-order non-linear differential equation for $y(\tau)$; making the change of variables $y(\tau) = \frac{1}{z(\tau)}$ results in the first-order linear differential equation
\begin{equation}
    -\tau z'(\tau)+\frac{1-\tau\sqrt{1-\frac{24\tilde{\xi}_3}{\tilde{c}_{\text{\tiny M}}}}}{\tau-1}z(\tau)+\frac{1}{2}\sqrt{\frac{\tilde{c}_{\text{\tiny M}}}{3}}=0\,,
\end{equation}
where the prime denotes the derivative with respect to $\tau$. The above equation may be solved for $z(\tau)$ by using the integrating factor method; substituting back to $y(\tau)$ gives
\begin{equation}
    y(\tau)=-2\beta_3\frac{\tau(1-\tau)^{\beta_3 -1}}{\sqrt{\frac{3}{\tilde{c}_{\text{\tiny M}}}}(1-\tau)^{\beta_3}-2\beta_3 c_1}\,.
\end{equation}
Combining $D_{\text{\tiny p}}(\tau)$ with $y(\tau)$ then yields the general solution
\begin{equation}
     D(\tau)=\frac{1}{\sqrt{3}(1-\tau)}\left(\sqrt{\tilde{c}_{\text{\tiny M}}}\left(\tau-2+\beta_3 \tau\right)+\frac{2\tau}{2c_1 (1-x)^{\beta_3} -\sqrt{\frac{3}{\tilde{c}_{\text{\tiny M}}}}\frac{1}{\beta_3}}\right)\,,
\end{equation}
where $\beta_3 =\sqrt{1-\frac{24\tilde{\xi}_3}{\tilde{c}_{\text{\tiny M}}}}$ and $c_1$ is an integration constant. Imposing the boundary condition $D'(\tau) = 0$, as discussed in the main text, fixes $c_1=\sqrt{\frac{3}{\tilde{c}_{\text{\tiny M}}}}\frac{1}{2\beta_3}$. Thus, the final solution is
\begin{equation}\label{Dsoln}
    D(\tau)=\sqrt{\frac{\tilde{c}_{\text{\tiny M}}}{3}}\frac{1}{\tau-1}\left(2-\tau+\beta_3 \tau\left(1-\frac{2}{1-(1-\tau)^{\beta_3}}\right)\right)\,.
\end{equation}

As discussed around equation \eqref{IsaddleHHLL} of the main text, evaluating the heavy-light vacuum $\bms_3$-block requires us to determine the integral $\int_0^t \dd\tau\frac{D(\tau)}{\tau}$. Rewriting $\frac{D(\tau)}{\tau}$ as
\begin{equation}\label{Dint}
    \frac{D(\tau)}{\tau}=\sqrt{\frac{\tilde{c}_{\text{\tiny M}}}{3}} \left(-\frac{2}{\tau} +\frac{\beta_3-1}{1-\tau}+\frac{2\beta_3(1-\tau)^{\beta_3-1}}{1-(1-\tau)^{\beta_3}}\right)\,,
\end{equation}
the integration results in 
\begin{equation}\label{D-int}
    \int\limits_0^t \dd\tau \frac{D(\tau)}{\tau} = \sqrt{\frac{\tilde{c}_{\text{\tiny M}}}{3}}\left(2\ln(1-(1-t)^{\beta_3})-(\beta_3 -1)\ln(1-t)-2\ln(t)\right).
\end{equation}

%%%%%%%%%%%%%%%%%%%%%%%
%%%%%%%%%%%%%%%%%%%

\subsection{Solving \texorpdfstring{$C(\tau)$}{C(t)}}
Plugging the solution \eqref{Dsoln} into equation \eqref{eq:C_diff} gives the equation for $C(\tau)$,
\begin{equation}\label{CDiffeq}
    C'(\tau)+f(\tau)C(\tau)+g(\tau)=0\,,
\end{equation}
where the functions $f(\tau)$ and $g(\tau)$ are given by
\begin{subequations}
\begin{align}
    f(\tau)&=\frac{1}{\tau}-\frac{1}{\tau(1-\tau)}\left(2+\tau\left(\beta_3-1+\frac{2\beta_3}{(1-\tau)^{\beta_3}-1}\right)\right)\,,\\
    \begin{split}
        g(\tau)&=\sqrt{\frac{3}{\tilde{c}_{\text{\tiny M}}}}\frac{4 \tilde{\Delta}_3 \tau^2}{(\tau-1)^2}-\frac{\tilde{c}_\text{\tiny L}}{\sqrt{3\tilde{c}_{\text{\tiny M}}}}\left(\frac{\tau\left(1+\beta_3-(1-\beta_3)(1-\tau)^{\beta_3}\right)+\left((1-\tau)^{\beta_3}-1\right)}{(\tau-1)^2 \left((1-\tau)^{\beta_3}-1\right)}\right)\\
        &\quad+\frac{\tilde{c}_{\text{\tiny L}} \tau^2}{2\sqrt{3\tilde{c}_{\text{\tiny M}}}}\left(\frac{\beta_3\left((1-\tau)^{2\beta_3}-1\right)-\left((1-\tau)^{\beta_3}-1\right)^2 -2\beta_3^2(1-\tau)^{\beta_3}}{(\tau-1)^2 \left((1-\tau)^{\beta_3}-1\right)^2}\right)\,.
    \end{split}
\end{align}
\end{subequations}
Equation \eqref{CDiffeq} is a first-order linear differential equation and it can be solved by using the integrating factor method; in particular the solution to $C(\tau)$ is given by the integral
\begin{equation}
\label{C:int}
    C(\tau)=-\frac{1}{\gamma(\tau)}\int\limits_{0}^\tau\!\! \dd\zeta\ \gamma(\zeta)g(\zeta)\,,
\end{equation}
where $\gamma(\tau)=\exp(\int_{0}^{\tau}\!\dd\zeta \, f(\zeta))$ is the integrating factor.  The integral in \eqref{C:int} can be performed using \emph{Mathematica}, which determines $C(\tau)$ up to a constant of integration $c_2$,
\begin{align}
\begin{split}
    C(\tau)=\frac{\tau(1-\tau)^{\beta_3 -1}}{\left((1-\tau)^{\beta_3}-1\right)^2}\Bigg[ & c_2 +\frac{(1-\tau)^{-\beta_3}}{2\sqrt{3\tilde{c}_{\text{\tiny M}}}\tau\beta_3}\bigg(\tilde{c}_{\text{\tiny L}} \tau\left(1+\beta_3 +(\beta_3 -1)(1-\tau)^{2\beta_3}\right)\\
    & +24\tilde{\Delta}_3 \tau\left((1-\tau)^{2\beta_3}-1\right)-2\tilde{c}_{\text{\tiny L}} \beta_3\left((1-\tau)^{\beta_3}-1\right)^2\\
    & -2\beta_3 \tau(1-\tau)^{\beta_3}\left(24\tilde{\Delta}_3 +\tilde{c}_{\text{\tiny L}}(\beta_{3}^{2}-1)\right)\ln(1-\tau))\bigg)\Bigg]\,.
\end{split}
\end{align}
Imposing the boundary condition $C'(0)=0$ fixes $c_2=- \frac{\tilde{c}_{\text{\tiny L}}}{\sqrt{3 \tilde{c}_{\text{\tiny M}}}}$. The solution for $C(\tau)$ is thus
\begin{align}
\begin{split}
    C(\tau)=&-\frac{\tau(1-\tau)^{\beta_3-1}\left(24\tilde{\Delta}_3+\tilde{c}_{\text{\tiny L}}(\beta_3^2-1)\right)}{\sqrt{3\tilde{c}_{\text{\tiny M}}}\left((1-\tau)^{\beta_3}-1\right)^2}\ln(1-\tau)\\
    & -\frac{12\tilde{\Delta}_3 \tau}{\sqrt{3\tilde{c}_{\text{\tiny M}}}\beta_3}\left(\frac{(1-\tau)^{\beta_3}+1}{(\tau-1)\left((1-\tau)^{\beta_3}-1\right)}\right)\\
    & -\frac{\tilde{c}_{\text{\tiny L}}}{2\sqrt{3\tilde{c}_{\text{\tiny M}}}(\tau-1)}\left(\tau-2-\frac{\tau}{\beta_3}\frac{(1-\tau)^{\beta_3}+1}{\left((1-\tau)^{\beta_3}-1\right)}\right)\,.
\end{split}
\end{align}

In the same way as for $D(\tau)$, evaluating the heavy-light vacuum $\bms_3$-block requires us to determine the integral $\int_0^t \dd\tau\frac{C(\tau)}{\tau}$. This integral can be solved using partial integration and the partial fraction expansion
\begin{equation}
    \frac{1+(1+\tau)^{\beta_3}}{(1-\tau)(1-(1-\tau)^{\beta_3}}=\frac{1}{(1-\tau)}+\frac{2(1+\tau)^{\beta_3-1}}{1-(1-\tau)^{\beta_3}}\,,
\end{equation}
which simplifies the integrand. The integral is then given by
\begin{align}\label{C-int}
\begin{split}
    &\int\limits_0^t  \dd\tau \frac{C(\tau)}{\tau} = \frac{ \tilde{c}_{\text{\tiny L}}}{\sqrt{3\tilde{c}_{\text{\tiny M}}}}\left(\ln(1-(1-t)^{\beta_3})-\ln(t)\right)+\frac{24}{\sqrt{3\tilde{c}_\text{\tiny M}}\beta_3^2}\left(\tilde{\Delta}_3-\frac{\tilde{c}_\text{\tiny L}}{\tilde{c}_\text{\tiny M}}\tilde{\xi}_3\right)\\
    &+\frac{1}{2\sqrt{3\tilde{c}_{\text{\tiny M}}}\beta_3}\left(\frac{24\tilde{\Delta}_3+\tilde{c}_{\text{\tiny L}}(\beta_3 -1)+(1-t)^{\beta_3}\left(24\tilde{\Delta}_3 +\tilde{c}_{\text{\tiny L}}(\beta_3 -1)(1+2\beta_3 )\right)}{1-(1-t)^{\beta_3}}\right)\ln(1-t)\,.
\end{split}
\end{align}
The constant term on the first line of the above solution contributes to the heavy-light vacuum $\bms_3$-block \eqref{HHLLsoln} as constant factor 
\begin{equation}
    \exp[\frac{24\xi_1}{c_\text{\tiny M}\beta_3^2}\left(\Delta_3-\frac{c_{\text{\tiny L}}}{c_{\text{\tiny M}}}\xi_3\right)]\,;
\end{equation}
such factors are of no consequence and may be neglected. 
\bibliographystyle{JHEP}
\bibliography{refs}
\end{document}